\documentclass{book}
\usepackage{amsmath}
\usepackage{amstext}
\usepackage{amssymb}
\usepackage{science}
\usepackage{multirow, graphicx}

\def\freeline[#1,#2,#3,#4]{\qbezier(#1,#2)(#1,#2)(#3,#4)}

\newcommand{\dummy}{\rule{0in}{0in}}

\newcommand{\Pl}{\mathrm{Pl}} 

\newcommand{\BI}{\beta} 
\newcommand{\Hil}{\mathcal{H}}

\newcommand{\hamp}{\mathcal{A}}
\newcommand{\tamp}{\mathcal{A}}
\newcommand{\scalaramp}{\tamp^{\text{scalar}}}
\newcommand{\gravamp}{\tamp^{\text{grav}}}
\newcommand{\hamplqg}{\hamp^{\text{LQG}}}

\newcommand{\tamplqg}{\tamp^{\text{LQG}}}

\newcommand{\spp}[1]{\mathbf{#1}}

\newcommand{\link}{\ell}
\newcommand{\node}{\nu}

\newcommand{\dual}{{}^\star}
\newcommand{\half}{\frac{1}{2}}

\newcommand{\pls}{\textbf{($\mathbf{+}$)}}
\newcommand{\negs}{\textbf{($\mathbf{-}$)}}
\newcommand{\degs}{\textbf{(deg)}}
\newcommand{\pns}{\textbf{($\mathbf{\pm}$)}}

\newcommand{\Regge}{\mathrm{R}}

\begin{document}

\renewcommand*{\thefootnote}{\fnsymbol{footnote}}
\renewcommand{\thesection}{\arabic{section}}
\renewcommand{\thesubsection}{\arabic{section}.\arabic{subsection}}

\chapter*{Spin Foams\footnote{To 
appear as a chapter of ``The Springer Handbook of Spacetime,'' edited by A. Ashtekar and V. Petkov
(Springer-Verlag, at Press)}}

\vspace{-0.5cm}

\noindent\textit{\Large Jonathan Engle}

\vspace{1cm}

Ever since special relativity, space and time have become  seamlessly
merged into a single entity, and space-time symmetries, such as Lorentz
invariance, have played a key role in our fundamental understanding of
nature. Quantum mechanics, however, did not originally conform to this new
way of thinking. The original formulation of quantum mechanics, called
`canonical', involves wavefunctions, operators, Hamiltonians, and time
evolution in a way that treats time very differently from space.  This
situation was improved by Feynman, who formulated quantum mechanics in
terms of probabilities calculated by summing over amplitudes associated to
classical histories --- the path integral formulation of quantum
mechanics.  As histories are naturally space-time objects in which space
and time can be viewed `on equal footing', the path integral formulation
allowed, for the first time, space-time symmetries to be manifest in a
general quantum theory.

The key insight of Einstein's theory of gravity, general relativity, is
that gravity is space-time geometry.  Space-time geometry, the one
`background structure' --- i.e., non-dynamical space-time structure ---
remaining after special relativity, was discovered to be dynamical and to
describe the gravitational field, revealing nature to be `background
independent.'  Background independence can equivalently be expressed in
terms of a profound enlargement of the basic space-time symmetry group of
physics: invariance under Lorentz transformations and translations is
replaced by invariance under the much larger group of \textit{space-time
diffeomorphisms}.

We have already seen in the chapter by Sahlmann on gravity, geometry and
the quantum, a canonical quantization of Einstein's gravity, and hence of
geometry, in which geometric operators are derived with discrete
eigenvalues \cite{al2004, rovelli2004, thiemann2007}.
Instead of space being a smooth continuum, we see that it comes in
discrete quanta --- minimal `chunks of space.'
Furthermore, as discussed in the chapter by Agullo and Corichi, when
applied to cosmology, this quantum theory of gravity leads to a new
understanding of the Big Bang in which usually problematic infinities are
resolved, and one can actually ask what happened \textit{before} the Big
Bang.  In spite of these successes, because it is a canonical theory, it
has as a drawback that space-time symmetries, in particular space-time
diffeomorphism symmetry, are not manifest. Equivalently, the preferred
separation between space and time prevents full background independence
from being manifest.

One can ask: Is there a way to construct a path-integral formulation of
quantum gravity, in which the most radical discovery of
general relativity, background independence, or equivalently, space-time
diffeomorphism invariance, can be fully manifest, which nevertheless
retains the successes of the canonical theory?
This is the question leading to the spin foam program.
In answering it one must understand more carefully
the relationship between the canonical and path integral formulations of
quantum mechanics, and in particular how these apply to
general relativity, with its special subtleties such as the `problem of
time' discussed in chapters by Claus Kiefer and Carlo Rovelli.
The end result is a path integral in which, instead of summing over
classical space-time histories, one sums over
\textit{histories of quantum states of space}.  These histories have a
natural space-time interpretation and thus
may be thought of as \textit{`quantum space-times'}. The resulting sum
over histories then provides
a framework for defining the \textit{dynamics} of loop quantum gravity in
which space and time are unified, in
the spirit of special and general relativity.
Due to their structure and the way they are labeled, these `quantum
space-times' have been named `spin foams' by John Baez \cite{baez1997}, a
name which thenceforth has been used to refer to the entire program.

In this chapter we hope to give the reader a broad view of the conceptual
ideas behind spin foams, the ideas that have led to
the spin foam model currently most often used in the community, as well as
provide a view of current avenues of investigation.
For a more detailed, complete review of spin foams, the recent reference
\cite{perez2012} is recommended to the interested reader.

\section{Background ideas}
\label{bkgr_sect}

\subsection{The path integral as a sum over histories of quantum states}
\label{qhist_sect}

The first formulation of quantum mechanics that was discovered, and that one learns, 
is the canonical formulation.
We review here briefly the basic structure of a canonical quantum theory.
The possible states of a canonical quantum system form a \textit{vector space}, that is,
they are such that states can be rescaled by real numbers and added to each other.
Additionally, one has an `inner product', which assigns to every two states $\phi$ and $\psi$ a complex number $\langle \psi, \phi \rangle$,
which may be roughly thought of as the `overlap' between states $\phi$ and $\psi$.
A vector space equipped with such an inner product is called a \textit{Hilbert space}; one often uses the phrase
``the Hilbert space of quantum states.''
For each possible measurable quantity, such as position, momentum, angular momentum, or energy --- or in the case of general relativity, areas of surfaces and volumes of regions ---  there is a corresponding operator $\hat{O}$ mapping states to states.
A number $\lambda$ is a
possible outcome of a measurement of $\hat{O}$ only if there exists a state $\psi$, such
that $\hat{O} \psi = \lambda \psi$. When the state of the system is $\psi$, then a measurement of $\hat{O}$ yields
$\lambda$ with certainty.   Such a $\lambda$ and corresponding $\psi$ are called an \textit{eigenvalue} and \textit{eigenstate}
of $\hat{O}$.  The set of all possible eigenvalues --- and hence possible results of a measurement --- of $\hat{O}$
is called the \textit{spectrum} of $\hat{O}$. Depending on the operator, its spectrum  may include all real numbers, or it may only include a discrete set of possible numbers.  This is the source of the name `quantum': that some quantities, when measured, can only come in discrete increments, called quanta.

Time evolution in canonical quantum theory is determined by Schr\"odinger's equation,
\begin{equation}
\label{schrodinger}
i \hbar \frac{d\psi}{dt} = \hat{H} \psi,
\end{equation}
where $\hbar$ is Planck's constant divided by $2\pi$, and $\hat{H}$ is the \textit{Hamiltonian} operator, which corresponds
to the total energy of the system.
If the system starts in an initial state $\psi(t_i)$, Sch\"odinger's equation
will uniquely determine its state $\psi(t_f)$ at any later time $t_f=t_i + T$,
thus providing a map $U(T)$ from possible initial states
$\psi(t_i)$ to final states $\psi(t_f)$, called a \textit{time evolution map}.
Using the time evolution map, and given two states $\psi_i, \psi_f$, and two times $t_i, t_f$, one can define a quantity
\begin{displaymath}
\tamp(\psi_f, t_f; \psi_i, t_i) := \langle \psi_f, U(t_f - t_i) \psi_i \rangle
\end{displaymath}
called a `\textit{transition amplitude}'.  The transition amplitude is of direct use for making predictions: If the system is prepared in an initial state $\psi_i$ at time $t_i$, the transition amplitude tells us the probability of measuring the system to be in a final state $\psi_f$
at time $t_f$. (Specifically, this probability is given by the formula
$|\tamp(\psi_f, t_f; \psi_i, t_i)|^2/|\langle \psi_f, \psi_f \rangle \langle \psi_i, \psi_i \rangle|$.)

The transition amplitude contains all information about the dynamics of the quantum system.
At the heart of the path integral formulation of quantum mechanics is Feynman's insight that the
transition amplitude can be \textit{rewritten} in terms of purely classical, \textit{space-time}
quantities.
Consider, for example, a single free particle, and consider the case in which $\psi_i$ and $\psi_f$ are `eigenstates of position',
i.e., states in which the position of the particle is exactly defined, being equal to some $x_i$ and $x_f$,
respectively.
We write $\psi_i = |x_i \rangle$ and $\psi_f = |x_f \rangle$.
In this case, one usually uses a simpler notation for the transition amplitude:
$\tamp(x_f, t_f; x_i, t_i) := \tamp\big(|x_f\rangle, t_i; |x_i\rangle, t_i\big)$.
The expression for the transition amplitude can be rewritten as
%
%
\newcommand{\UTN}{U{\textstyle\left(\frac{T}{N}\right)}}
\begin{equation}
\label{steptwo}
\tamp(x_f, t_f; x_i, t_i)
= \langle x_f, \UTN \cdots \UTN \UTN \UTN x_i \rangle
%
%
\end{equation}
where $T:= t_f - t_i$ and one has used the fact that the time evolution $U(T)$ is equivalent to performing $N$ evolutions over the
smaller time $T/N$.  The eigenstates of position $|x\rangle$ satisfy the following identity:
For all $\psi, \phi \in \Hil$,
\begin{equation}
\label{res}
\langle \psi, \phi \rangle = \int_{-\infty}^\infty \langle \psi, x \rangle \langle x, \phi \rangle dx .
\end{equation}
This is known as a `completeness relation' or `resolution of the identity'.  Note that the range of integration on the right hand side
includes all possible values which can result from a measurement of the position $\hat{x}$ --- that is, the integral is over the spectrum of $\hat{x}$.
If $\hat{x}$ were `quantized', that is, if its spectrum were discrete, this integral would be replaced by a \textit{sum} over the discrete spectrum.
We will remark on this later.
Applying the identity (\ref{res}) to (\ref{steptwo}) $N-1$ times, in sequence, one obtains
\begin{eqnarray*}
&& \hspace{-0.8cm} \tamp(x_f, t_f; x_i, t_i)
= \int \langle x_f, \UTN \cdots \UTN \UTN x_1 \rangle \langle x_1, \UTN x_i \rangle dx_1  \\
&&
= \int \int
\langle x_f, \UTN \cdots \UTN x_2 \rangle \langle x_2, \UTN x_1 \rangle \langle x_1, \UTN x_i \rangle  dx_1  dx_2 \\
&& \hspace{0.7cm} \vdots \\
&&
= \int  \int  \cdots \int
\langle x_f, \UTN x_{N-1} \rangle \cdots \langle x_2, \UTN x_1 \rangle \langle x_1, \UTN x_i \rangle dx_1 dx_2 \cdots dx_{N-1} .
\end{eqnarray*}
In this expression one has introduced $N-1$ intermediate position eigenstates, and one integrates over all possible such intermediate
states.  This sequence of intermediate states forms a discrete history of quantum states.
Note that the above expression is exact for any $N$.
If one takes the limit at $N$ approaches infinity, the discrete histories are replaced by continuum histories of
quantum states, and one obtains the \textit{path integral};
\begin{equation}
\label{pathint}
\tamp(x_f, t_f; x_i, t_i) = \int_{\substack{x(t_i) = x_i \\ x(t_f) = x_f}}
\exp\left(\frac{i}{\hbar}S[x(\cdot)]\right) \mathcal{D} x(\cdot)
\end{equation}
where, heuristically, $\mathcal{D} x(\cdot)$ denotes `$\prod_{t} dx(t)$', and
$S[x(\cdot)]$ is the \textit{action} for the theory.  The action is a purely classical quantity, which specifies a number
for each possible classical history $x(t)$. It is maximized or minimized when $x(t)$ is a \textit{solution to the classical
equations of motion}.
Because of the close relation between integrals and sums (one is just a limit of the other),
the integral in equation (\ref{pathint}) is also loosely referred to as a \textit{sum} over paths, or a \textit{sum} over histories.
If the position operator $\hat{x}$ had had a \textit{discrete} spectrum, so that only a discrete set of values were allowed for $x$,
as already mentioned, the resolution of the identity (\ref{res}) would have actually been replaced by a sum, and the final path integral
(\ref{pathint}) would have actually become a sum rather than an integral.  There are also cases where the final expression for the
propagator (\ref{pathint}) involves a combination of sums and integrals. In this chapter, as in much of the literature on
path integrals, we will be loose with the distinction between sums over paths and integrals over paths, and will use the terms
``path integral'', ``sum over paths'', and ``sum over histories'' interchangeably. Nevertheless, because, in the case of quantum gravity,
the primary interest of this chapter, one will turn out to have mostly sums, we will generally prefer to use the term sum.

Equation (\ref{pathint}) provides an expression for the transition amplitude from a position $x_i$ at time $t_i$
to a position $x_f$ at time $t_f$, an expression that involves only a sum over \textit{classical paths} $x(t)$ that start at $x_i$ and end at $x_f$,
and the \textit{classical} action $S[x(t)]$ depending on this path.
The canonical theory enters into the expression in \textit{one way only}:
It determines the spectrum of $\hat{x}$ and hence the allowed
values that the history of eigenvalues $x(t)$ can take at each moment in time.
Other than this, \textit{classical} physics is the only input for this expression. Because of this, Feynman made the radical proposal that this formula, which encodes all physical predictions for the system in question, be a new starting point for the very definition of the quantum theory.

However, care is necessary. As noted, one piece of information from the canonical quantum theory \textit{does} remain: It is the \textit{canonical} theory which tells us the spectrum of the position operator $\hat{x}$, and hence the possible positions which one
sums over in the path integral.  In the case of the free particle,
the spectrum of the position includes all real numbers, so that, in fact, the sum is equivalent to a sum over all
classical histories.  However, in other theories this is not necessarily the case.  In particular, in the case of gravity, one must sum over \textit{histories of geometry}.  But, one of the seminal results of loop quantum gravity is that
\textit{geometry is quantized}.  Areas of surfaces and volumes of regions can only take on discrete sets of possible values.
Thus, one should not sum over all histories of
\textit{classical} geometries, but rather over histories of the allowable \textit{quantum geometries} predicted by loop quantum gravity.
This is the insight leading to the spin foam program.

Before closing this section, let us remark that the integrand in equation (\ref{pathint}) can be interpreted as giving the probability amplitude for a
\textit{single history} $x(t)$:
\begin{equation}
\label{histamp}
\hamp[x(\cdot)] = \exp \left(\frac{i}{\hbar}S[x(\cdot)]\right) .
\end{equation}
The total transition amplitude (\ref{pathint}) is then obtained by integrating (or adding) the amplitudes (\ref{histamp}) associated to
all histories compatible with with the relevant `boundary conditions', $x(t_i) = x_i$, $x(t_f) = x_f$.
%
%
%

\begin{figure}
\begin{center}
\dummy\hspace{0.6in}
\begin{minipage}{1.8in}
\setlength{\unitlength}{2in}
\begin{picture}(1,1.7)
\includegraphics[height=3.53in]{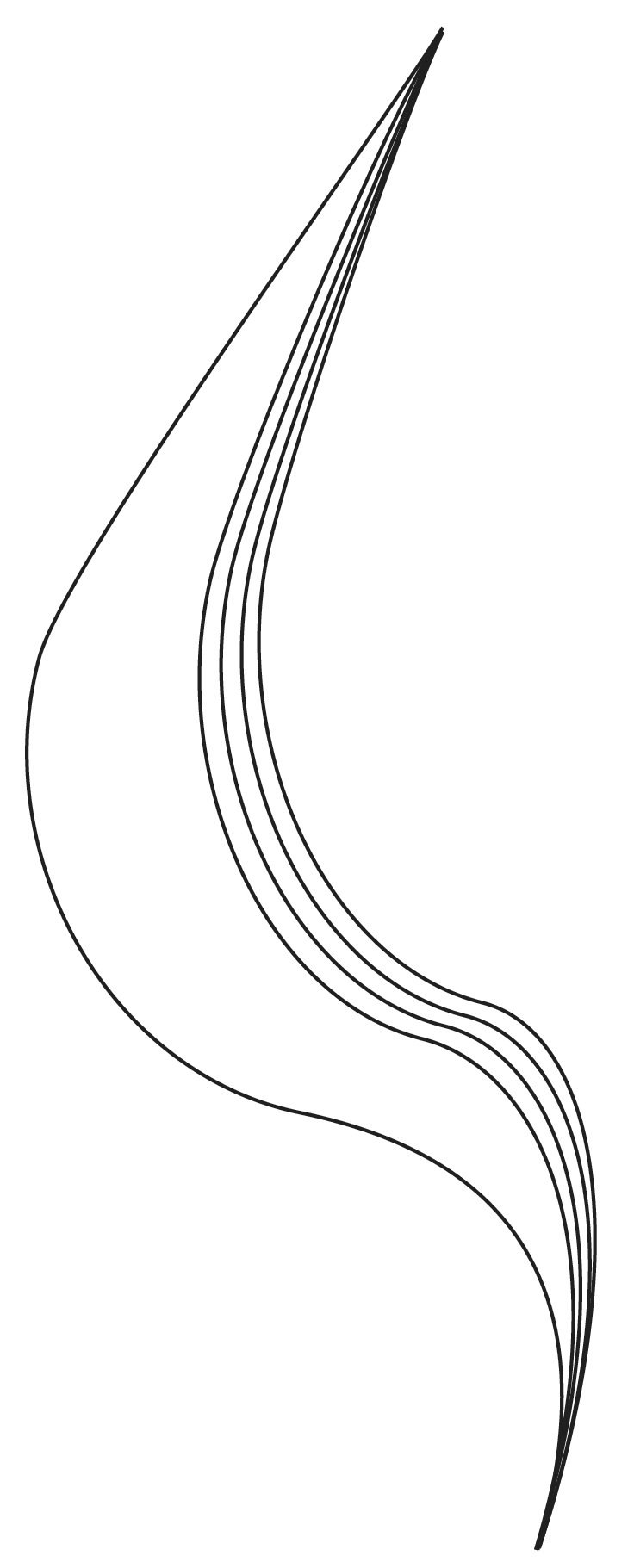}
\put(-0.835,0.8){$x_1(t)$}
\put(-0.618,0.8){$x_2(t)$}
\put(-0.335,0.8){$\dots$}
\end{picture}
\end{minipage}
\hspace{0.4in}
\begin{minipage}{1.8in}
\setlength{\unitlength}{2in}
\begin{picture}(0.9,0.9)
\includegraphics[height=1.2in]{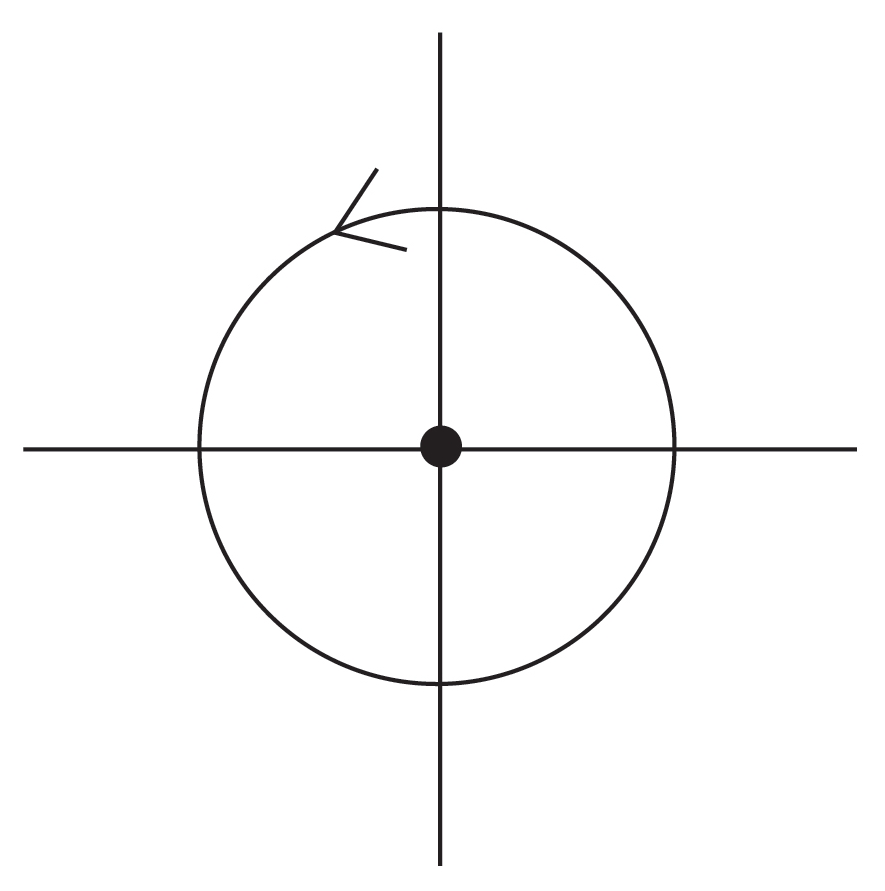}
\put(-0.285,0.31){$0$}
\put(-0.1,0.31){$\mathrm{Re}z$}
\put(-0.29,0.505){$\mathrm{Im}z$}
\put(-0.75,-0.13){
$e^{iS[x_1(\cdot)]/\hbar} + e^{iS[x_2(\cdot)]/\hbar}+ \dots
\sim 0$}
\end{picture}\\
\vspace{0.3in}

\dummy
\end{minipage}
\end{center}

\caption{\label{sclim_fig}
In the classical limit, at histories $x(\cdot)$ where $\frac{1}{\hbar}S[x(\cdot)]$ changes, it changes very fast, so that the phases from the
sum over histories near $x(\cdot)$ tend to cancel (illustrated in figure).
When $S[x(\cdot)]$ does \textit{not} change with $x(\cdot)$, the phases do not cancel, but reinforce each other.  This happens
when $x(\cdot)$ is a local minimum or maximum of $S[x(\cdot)]$ --- that is, when $x(\cdot)$ is a solution to the classical equations of motion.
In this way the classical solutions dominate the sum over histories in the classical limit.
}
\end{figure}
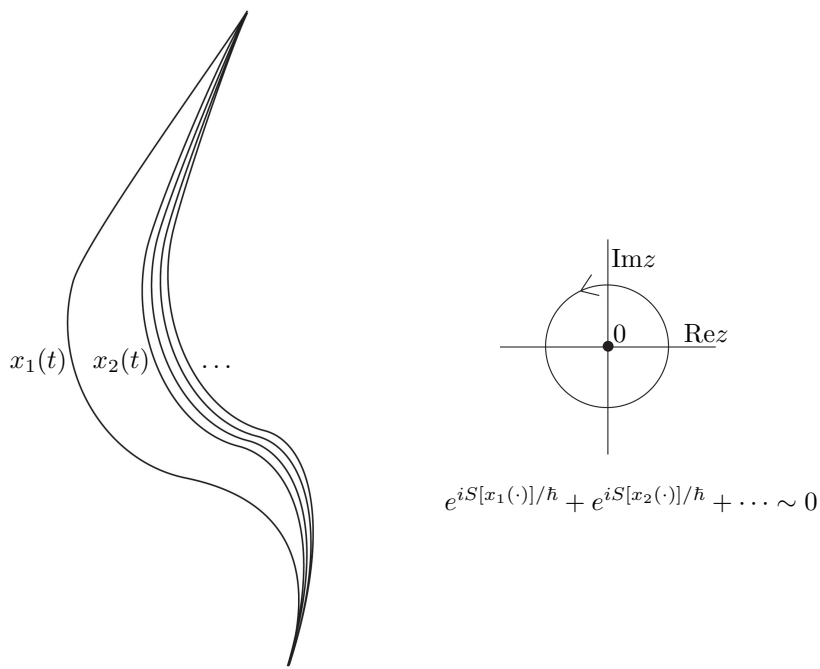

The precise form (\ref{histamp}) for the amplitude of each history not only  arises from the canonical quantum theory
in the manner presented above,  but it is also important for the correct \textit{classical limit} of
the quantum theory.
When constructing a quantum theory, usually the corresponding classical theory is already well-tested experimentally.
In order to be consistent with known experiments, it is therefore crucial that the predictions of the quantum theory
agree with those of the classical theory in situations where the effects of quantum mechanics can be neglected.
One way of stating this requirement is that if appropriate combinations of the physical scales in the situation are large compared to
Planck's constant (so that Planck's constant can effectively be scaled to zero) then the quantum theory should yield
the same predictions as the corresponding classical theory.
The limit here described --- that of either large physical scales or Planck's constant being scaled to zero ---
is what is called the `classical limit' of a quantum theory, and the requirement that this yield predictions equivalent to the
classical theory is called the requirement of having the \textit{correct classical limit}.
Let us consider the classical limit of the path integral.  For the present argument, it is easiest to cast this as the limit in which
\textit{Planck's constant is scaled to zero}.  In this limit, the phase $\frac{1}{\hbar}S$ of the amplitude (\ref{histamp})
becomes very large compared to $2\pi$. If one divides up the domain of integration --- the space of histories compatible with the boundary conditions
--- into many small neighborhoods, one finds that, in the vast majority of these neighborhoods, the phase of the integrand
will oscillate very fast.  As a consequence, in such neighborhoods, there tend to be an equal number of opposite phase contributions
from the path integral which cancel each other, so that the total contribution from such neighborhoods tends to be zero. (See figure \ref{sclim_fig}.)
The only neighborhoods where the phase is \textit{not} oscillating fast are those where $S[x(\cdot)]$ does not change very much
when $x(\cdot)$ changes.  These are precisely the neighborhoods where $S[x(\cdot)]$ is \textit{maximum} or \textit{minimum}, that is,
precisely the neighborhoods containing a \textit{solution to the classical equations of motion}.
Thus, one sees that, in the classical limit, 
only histories near solutions to the classical equations of motion contribute to the path integral.  
This is key to obtaining the correct classical limit of the quantum theory.  
The Feynman prescription (\ref{histamp}) 
for the probability amplitude of a single history is thus directly related to ensuring  the correct classical limit.

\subsection{Field theory and the general boundary formulation of quantum mechanics}

Before going on to the specific case of gravity, we take the opportunity to first discuss field theory,
and introduce what is known as the \textit{general boundary} formulation of quantum mechanics
\cite{oeckl2003, cdort2003, rovelli2004}.
In the case of field theory, instead of integrating over possible paths $x(t)$ of a particle from
time $t=t_i$ to $t=t_f$ as in (\ref{pathint}), one integrates over possible \textit{fields} $\phi(x)$ on the
\textit{four dimensional space-time region} bounded by the instants $t=t_i$ and $t=t_f$.
In the general boundary formulation, this region is allowed to be replaced by \textit{any} space-time region.
The biggest advantage of this formulation of quantum mechanics is that, by choosing this region to be finite, it permits
purely local calculations in a quantum field theory in which one need not worry about the asymptotic behavior
of states at infinity.
Not only are such calculations more consistent with the locality of the measuring apparatus one would actually
use, but they are technically simpler, and have been central to most work in spin foams up until now.

\subsubsection*{The free scalar field}

\begin{figure}
\begin{center}
\setlength{\unitlength}{1.3in}
\begin{picture}(1.2,1.5)
\linethickness{1.4mm}

\freeline[0,1.2,1.5,1.2]
\freeline[0,0.2,1.5,0.2]

\linethickness{0.2mm}

\freeline[-0.015,0.2,-0.015,1.2]
\freeline[1.515,0.2, 1.515,1.2]

\put(0.5,1.3){\Large $\Sigma_{t_f}, \varphi_f(\spp{x})$}
\put(0.55,0.65){\Large $R, \phi(x)$}
\put(0.5,0){\Large $\Sigma_{t_i}, \varphi_i(\spp{x})$}
\end{picture}
\end{center}
\caption{\label{Rphi_fig}
In the path integral for the scalar field, one sums over all \textit{fields} $\phi(x)$ on some \textit{space-time region} $R$
compatible with given initial values $\varphi_i(\spp{x})$ on the initial hypersuface $\Sigma_{t_i}$ and
final values $\varphi_f(\spp{x})$ on the final hypersurface $\Sigma_{t_f}$, where $\Sigma_{t_i}$ and $\Sigma_{t_f}$
bound $R$.
}
\end{figure}
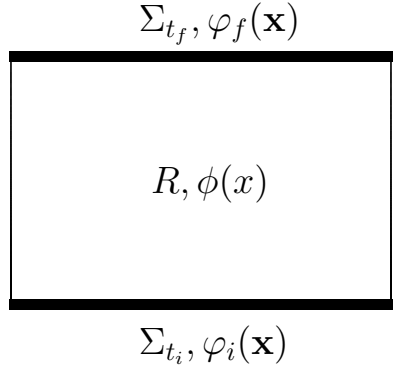

As an example, let us look at the case of a scalar field in Minkowksi space.
In this case, one has as basic canonical variables $\varphi(\spp{x})$ and its conjugate momentum
field $\pi(\spp{x})$, and corresponding operators $\hat{\varphi}(\spp{x}), \hat{\pi}(\spp{x})$.
We here use bold to denote spatial points.
One has a complete set of simultaneous eigenstates $|\varphi(\spp{x})\rangle$ of the
operators $\hat{\varphi}(\spp{x})$, each now labeled by a \textit{field} $\varphi(\spp{x})$ on space.
A history of such fields, $\phi(t, \spp{x}) = \phi(x)$ is a field on the four dimensional \textit{space-time region} $R$
bounded by the three dimensional ``instant time'' hypersurfaces $t=t_i$ and $t=t_f$, which shall be denoted
$\Sigma_{t_i}$ and $\Sigma_{t_f}$, respectively (see figure \ref{Rphi_fig}). Note that space-\textit{time} points such as $x$ will not be bolded.
Equation (\ref{pathint}) becomes, in this case,
\begin{equation}
\label{pathintphi}
\scalaramp(\varphi_f, t_f; \varphi_i, t_i) :=
\scalaramp(|\varphi_f \rangle, t_f ; |\varphi_i \rangle, t_i) = \int_{\substack{\phi|_{t_i} = \varphi_i\\ \phi|_{t_f} = \varphi_f}}
 e^{i S[\phi]} \mathcal{D} \phi
\end{equation}
where $S[\phi]$ is the classical action (the exact form is not important for the present discussion).
Next, note that the field $\varphi_i$ is a field on the hypersurface $\Sigma_{t_i}$, and $\varphi_f$ is a field on the
hypersurface $\Sigma_{t_f}$.
These two hypersurfaces together form the boundary of the four-dimensional space-time region $R$, the region on which
the field $\phi$ is defined. Let $\varphi$ denote the combination
of the fields $\varphi_i, \varphi_f$ on the \textit{full} boundary of $R$, denoted $\partial R$, which in this case is equal to $\Sigma_f \cup \Sigma_i$.
The state $|\varphi_i \rangle$ can be thought of as living in a copy $\Hil_{\Sigma_{t_i}}$ of the Hilbert space
associated to the surface $\Sigma_{t_i}$, and $|\varphi_f \rangle$ as living in a copy $\Hil_{\Sigma_{t_f}}$
of the Hilbert space of quantum states associated to the surface $\Sigma_{t_f}$.   The full field $\varphi$ on all of $\partial R$
can then be thought of as labelling a state $|\varphi \rangle$  in a certain \textit{combined} Hilbert space
$\Hil_{\partial R}$ for the \textit{full} boundary of $R$.

Let us define the Hilbert space $\Hil_{\partial R}$. Consider a given Hilbert space of quantum states $\Hil$.
Often one thinks of quantum states $|\Psi \rangle \in \Hil$ as `column vectors' (`kets').  Their Hermitian
conjugates, denoted $|\Psi \rangle^\dagger =: \langle \Psi |$ are then `row vectors' (`bras').  The inner product between
two states $\Psi, \Phi$ can then be written as the matrix product of the row vector $\langle \Psi |$ with the column vector
$|\Phi \rangle$, yielding a complex number:
\begin{displaymath}
\langle \Psi | \Phi \rangle = \langle \Psi, \Phi \rangle
\end{displaymath}
(whence the motivation for the notation $\langle \Psi |$ and $|\Phi \rangle$).
The space of `row vectors' is the called the space \textit{dual} to $\Hil$, and is written $\Hil^*$.
The Hilbert space $\Hil_{\partial R}$ for the \textit{full} boundary $\partial R = \Sigma_{t_f} \cup \Sigma_{t_i}$,
in terms of $\Hil_{\Sigma_{t_i}}$ and $\Hil_{\Sigma_{t_f}}$,
is then defined to consist in formal sums of products of states in the dual $\Hil_{\Sigma_{t_f}}^*$ and in $\Hil_{\Sigma_{t_i}}$
(the product is denoted using the symbol `$\otimes$'). Mathematically, this is expressed by saying that $\Hil_{\partial R}$ is the
\textit{tensor product} of $\Hil_{\Sigma_{t_f}}^*$ with $\Hil_{\Sigma_{t_i}}$, and one writes
$\Hil_{\partial R}:= \Hil_{\Sigma_{t_f}}^* \otimes \Hil_{\Sigma_{t_i}}$.
In terms of the initial and final field eigenstates $|\varphi_i\rangle \in \Hil_{\Sigma_{t_i}}, |\varphi_f\rangle\in \Hil_{\Sigma_{t_f}}$,
the corresponding field eigenstate on the \textit{full} boundary of $R$ is given by
$|\varphi \rangle := |\varphi_f\rangle^\dagger \otimes |\varphi_i \rangle = \langle \varphi_f|\otimes |\varphi_i \rangle
\in \Hil_{\partial R}$.
The Hilbert space $\Hil_{\partial R}$ on the full boundary of $R$ is called the \textit{boundary Hilbert space},
and $|\varphi \rangle$ is called a \textit{boundary state}.

\begin{figure}
\begin{center}
\setlength{\unitlength}{1in}
\begin{picture}(1.5,1.5)
\includegraphics[height=1.5in]{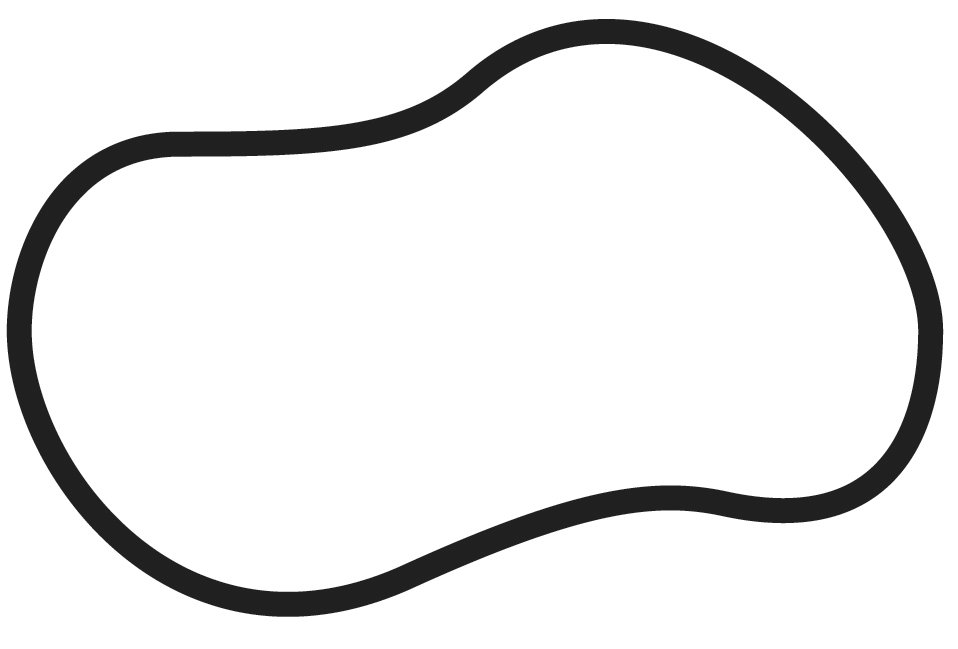}
\put(-1.4, 0.65){\Large $R, \phi(x)$}
\put(0.02,0.65){\Large $\partial R, \varphi(\spp{x})$}
\end{picture}
\hspace{1in}\dummy
\end{center}
\caption{\label{compactR}
The general boundary formulation of the path integral applies even when the space-time region $R$ is compact.
}
\end{figure}

In terms of the label $\varphi$ and \textit{boundary} states,
equation (\ref{pathintphi}) becomes
\begin{equation}
\label{pathintR}
\scalaramp(\varphi, R) \equiv \scalaramp(|\varphi\rangle, R) = \int_{\phi|_{\partial R} = \varphi}
e^{i S[\phi]} \mathcal{D} \phi.
\end{equation}
This expression has the benefit that it makes sense also when $R$ is \textit{any} space-time region,
leading to a natural generalization of the path integral formalism.  This generalization is called the
\textit{general boundary} formulation of quantum mechanics, and is
equivalent to the more standard formulations of quantum mechanics \cite{oeckl2003, cdort2003, rovelli2004}.
The interpretation of the path integral (\ref{pathintR}) is the direct generalization of the interpretation
of the original path integral (\ref{pathintphi}): It provides the probability amplitude
of measuring the field $\phi$ to have the values $\varphi$ on the boundary of the region $R$.
The expression (\ref{pathintR}) applies when the boundary state is an eigenstate $|\varphi \rangle$
of the scalar field operator $\hat{\varphi}(\spp{x})$; from this one can deduce the amplitude
$\scalaramp(\Psi, R)$  for \textit{any} quantum boundary state $\Psi$ in $\Hil_{\partial R}$.
The general boundary formalism applies even, and in our case most importantly,
when $R$ is compact. (See figure \ref{compactR}.)
One advantage of this generalized formalism when $R$ is chosen to be compact,
is that one can completely side step the issue of how the quantum state behaves as one
approaches spatial infinity, an issue which shouldn't matter for concrete applications anyway, because
one never measures fields at infinity in actual experiments.  Furthermore, the lack of an a priori fixed notion
of which space-time regions may be used is more consistent with the spirit of
 background independence which will be central in the case of quantum gravity.

\subsection{The case of gravity: The `problem of time' and the path integral as projector.}
\label{gengrav_sect}

Applying the above ideas to gravity involves unique subtleties.
Specifically, in general relativity, when initial data surfaces are compact and without boundary (so that there are no boundary terms),
the Hamiltonian $H$ is constrained to be \textit{zero}.  In fact, the Hamiltonian can be expressed in terms of a Hamiltonian
density $H = \int \mathcal{H}(\spp{x}) d^3 \spp{x}$, and this Hamiltonian density $\mathcal{H}(\spp{x})$
is constrained to be zero \textit{at each point} $\spp{x}$.
Because $\mathcal{H}(\spp{x})$ is constrained to be zero, it is called the Hamiltonian \textit{constraint}.
In the quantum theory, the Hamiltonian constraint dictates that states be eigenstates of the Hamiltonian constraint operator
$\hat{\mathcal{H}}(\spp{x})$ with eigenvalue zero --- that is, one requires
that states be annihilated by the Hamiltonian constraint, $\hat{\mathcal{H}}(\spp{x}) \Psi = 0$,
and hence, also by the Hamiltonian, $\hat{H} \Psi = 0$.
By Schr\"odinger's equation (\ref{schrodinger}), this implies the curious property
\begin{equation}
\label{WdW}
\frac{d\Psi}{dt} = \left(\frac{-i}{\hbar}\right) \hat{H} \Psi = 0,
\end{equation}
i.e., that the quantum state should not evolve in time. This fact is directly related to the background
independence of general relativity: that there is no background time variable. Whereas in \textit{classical} general relativity
one can introduce an arbitrary time variable for convenience, in \textit{quantum} general relativity, even introducing such a time
for convenience is forbidden, or at least useless.

It is clear, therefore, that in quantum gravity one cannot interpret
the Feynman path integral in terms of time evolution, as was done in (\ref{pathint}).  In fact, the interpretation is different.
Instead, in the interpretation of the path integral, the time evolution map is replaced by a \textit{projector} $P$ onto solutions
of $\hat{\mathcal{H}}(\spp{x}) \Psi = 0$, the quantum Hamiltonian constraint \cite{leutwyler1964, hh1983, rr1996}.
Let us be concrete.  In the case of gravity, the space-time field is the four dimensional \textit{metric},
denoted $g(x)$, which determines the lengths of, and angles between, vectors at each point $x$, which
in turn determines geometrical lengths of curves, areas of surfaces, volumes of regions etc. --- that is, $g(x)$ determines the geometry of space-time.
The \textit{canonical} variables on a given instant-time hypersurface $\Sigma_t$ are the \textit{three} dimensional metric $h(\spp{x})$
determining the three dimensional geometry of $\Sigma_t$,
and its conjugate momentum $\Pi(\spp{x})$,
which determines the way $\Sigma_t$ curves in the larger four dimensional space-time
and can be related to the time derivative of $h(\spp{x})$.
Hence, in the quantum theory one has operators $\hat{h}(\spp{x})$ and $\hat{\Pi}(\spp{x})$,
and simultaneous eigenstates $|h\rangle$ of the operators $\hat{h}(\spp{x})$.
The states $|h\rangle$ and the projector $P$ are then related to the Feynman path integral by
\begin{equation}
\label{projpathint}
\langle h_f, P \,\, h_i \rangle =  \int_{\substack{g|_{\Sigma_{t_i}} = h_i\\
g|_{\Sigma_{t_f}} = h_f}}
 e^{i S[g]} \mathcal{D} g
\end{equation}
where $g|_{\Sigma} = h$ means that the geometry
induced by $g$ on $\Sigma$ is equal to $h$.

Another way of stating this phenomenon is that equation 
(\ref{WdW}) is simply a statement of \textit{gauge invariance} of the wavefunction --- time
translations are coordinate transformations, and hence do not change the physical state, and so are gauge.  At the same
time, it is also a statement of the quantum version of the component
$H = \int \mathcal{H}(\spp{x}) d^3 \spp{x} = 0$ of the Hamiltonian constraint.  In fact, in
general, for every gauge symmetry in a system, there is a corresponding constraint, and, as happens here, in the quantum
theory, invariance under the gauge symmetry and satisfaction of the corresponding quantum constraint become
one and the same thing.
Constraints related to gauge in this way are called \textit{first class} \cite{dirac1964}.
Not only is $H$ a first class constraint, but so are the infinity of individual Hamiltonian constraints
$\mathcal{H}(\spp{x})=0$ for each point $\spp{x}$.
In fact, all other fields which mediate forces in nature (electroweak and strong forces) also have first class constraints
and corresponding gauge symmetries.
Quite generally, whenever a system has first class constraints,
the path integral projects onto solutions of the first class constraints, so that the projection property seen in (\ref{projpathint}) is not unique to general relativity \cite{leutwyler1964}.

Exactly as in the case of the scalar field theory in the last subsection,
the expression (\ref{projpathint}) generalizes to an arbitrary space-time region.
If $\gravamp(\Psi, R)$ denotes the probability amplitude for a given quantum gravity state $\Psi$ on the boundary of a given
region $R$, and $h$ denotes a given three-dimensional metric on the boundary $\partial R$ of $R$,
and $|h\rangle$ the corresponding eigenstate in the boundary state space, we have
\begin{equation}
\label{gampeq}
\gravamp(h, R) := \gravamp(|h\rangle, R)
= \int_{g|_{\partial R} = h} e^{iS[g]}  \mathcal{D} g .
\end{equation}

We close this section with a remark.
In the case of a scalar field, there is a \textit{background space-time geometry}, $\mathring{g}$,
%
%
present and the action $S[\phi]$ depends on it: $S[\phi] = S[\phi, \mathring{g}]$.
Because of this, $\scalaramp(\Psi, R)$ in fact depends
on the \textit{size} and \textit{shape} of the chosen region $R$, as determined by this background
geometry.   By contrast, in the case of quantum gravity, there is no background geometry, and so $R$
has no non-dynamically defined `shape' or `size'.
In this case the boundary quantum state $\Psi$ codes the information about geometry, which is now
dynamical.
If $\Psi$ is sufficiently peaked on a classical geometry, then $R$ again has a shape, but this
shape is determined by $\Psi$, and not by any background geometry.

\section{Spin foam models of quantum gravity}

\subsection{Review of spin network states and their meaning}
\label{sn_sect}

It is now time to incorporate into the discussion what has been learned from the canonical quantization of gravity
known as (canonical) loop quantum gravity (LQG). The Hilbert space of states in LQG is spanned by
what are called \textit{spin networks} (as discussed in the chapter by Sahlmann).
In this chapter, because it will be most useful later on,
we review a form of spin network introduced by Livine and Speziale \cite{ls2007},
which we will refer to as Livine-Speziale spin networks. (In the literature they are more
commonly referred to as ``Livine-Speziale coherent states''.)
%
%
Each such spin network state is `peaked' on a particular three-dimensional, discrete,
spatial geometry.  We first review how each spin network is labeled, and then how these labels
determine the corresponding geometry.

\begin{figure}
\begin{center}
\setlength{\unitlength}{1in}
\begin{picture}(2,1.7)
\includegraphics[height=1.7in]{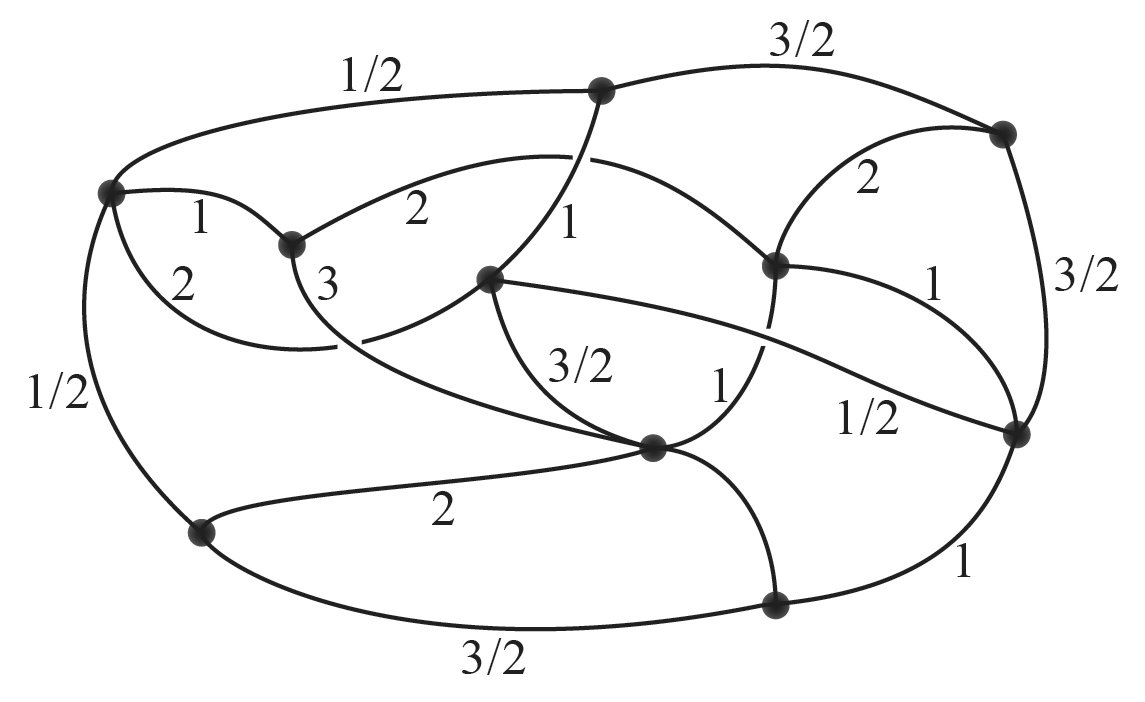}
\end{picture}
\end{center}
\caption{\label{spinnet_fig}
Each spin network state is labeled by a choice of graph, with spins labelling the links, and other quantum numbers labeling the nodes.
}
\end{figure}

\begin{figure}[t]
\begin{minipage}[t]{2.85in}
\centering
\dummy \hspace{-1.3in}
\setlength{\unitlength}{0.9412in}
\begin{picture}(1.7,1.7)
%
%
\includegraphics[height=1.6in]{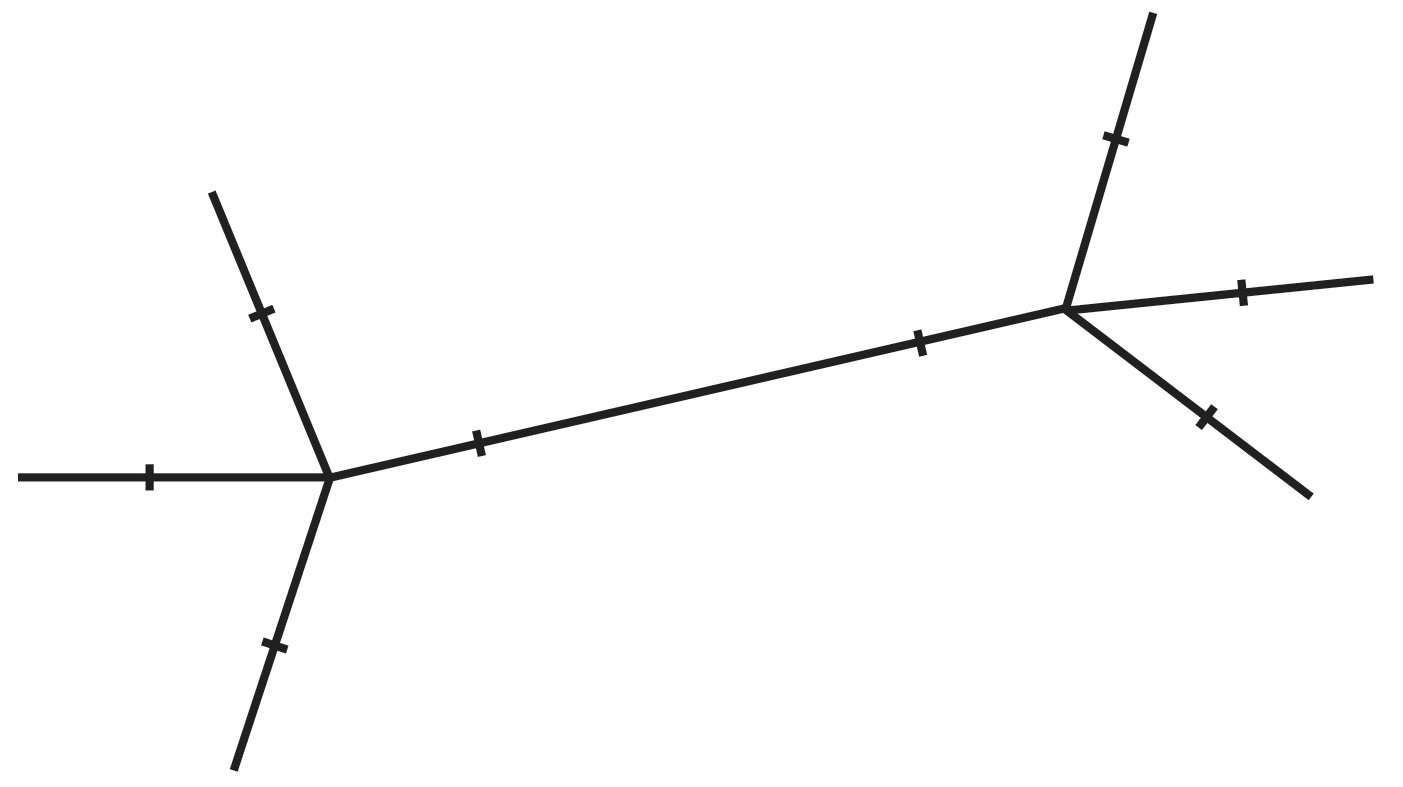}
\put(-1.6,0.9){$j_\link$}
\put(-2.2,0.6){$n_{\node\link}$}
\put(-2.4,0.9){$n_{\node\link_1}$}
\put(-2.65,0.73){$n_{\node\link_2}$}
\put(-2.37,0.35){$n_{\node\link_3}$}
\put(-1.045,1.055){$n_{\node'\link}$}
\put(-0.96,1.32){$n_{\node'\link_1'}$}
\put(-0.65,1.14){$n_{\node'\link_2'}$}
\put(-0.77,0.79){$n_{\node'\link_3'}$}
\end{picture}
\caption{\label{snlabels_fig}
Each link $\link$ is labeled by a spin $j_\link$.  For each node $n$, and each link $\link$ incident at $n$,
one has also a unit three dimensional vector $n_{\node\link}$.
}
\end{minipage}
\quad
\begin{minipage}[t]{2.85in}
\centering
\setlength{\unitlength}{1in}
\begin{picture}(1.7,1.7)
\includegraphics[height=1.7in]{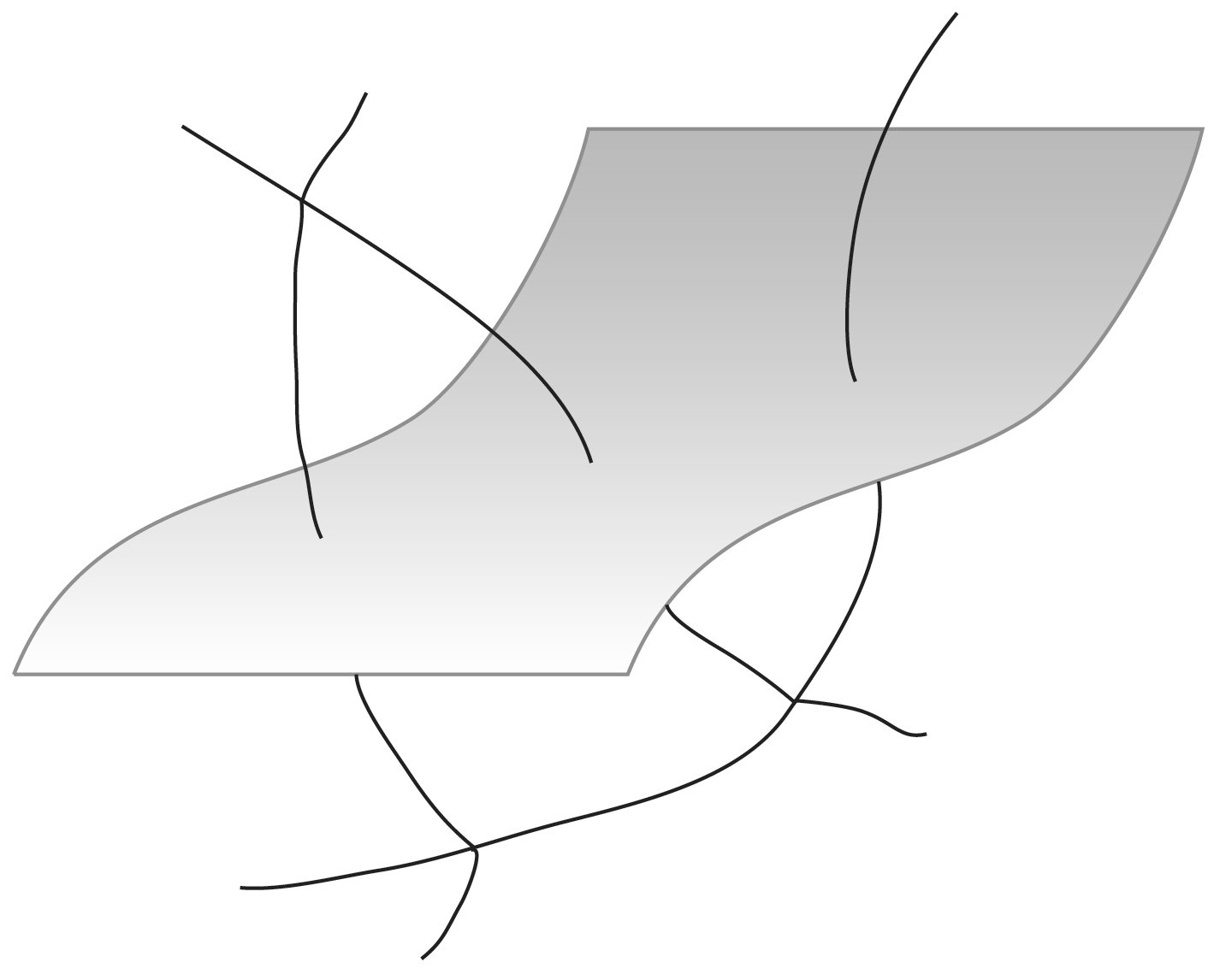}
\put(-1.5,0.8){$j_1$}
\put(-1.08,1){$j_2$}
\put(-0.78,1.2){$j_3$}
\end{picture}
\caption{\label{qarea} Each spin network link, with spin $j_\link$, intersecting a surface $S$ contributes to the surface an area of
$8\pi \ell_\Pl^2 \BI \sqrt{j_\link (j_\link + 1)}$, where $\ell_\Pl$ is the Planck length.}
\end{minipage}
\end{figure}

Each spin network state is first labeled by a collection of curves in space which intersect each other at
most at their end points.  Such a collection of curves is called a \textit{graph} and will be typically denoted $\gamma$.
(See figure \ref{spinnet_fig}.)
Following the terminology of Rovelli \cite{rovelli2004}, we call each curve in the graph
a \textit{link}, and each endpoint of a curve a \textit{node}.  Each link $\link$
is labeled by a half integer spin $j_\link = 0, \frac{1}{2}, 1, \frac{3}{2}, \dots$.
At each node $\node$, and for each link $\link$ ending or beginning at $\node$, there is furthermore
a unit, three-dimensional vector $n_{\node\link}$. (See figure \ref{snlabels_fig}.)
We write $|\gamma, \{j_\link, n_{\node\link}\} \rangle$ to denote such a spin network.

The labels $\gamma, \{j_\link, n_{\node\link}\}$ determine the spatial geometry by determining areas of
surfaces and volumes of regions.
In determining these areas and volumes, an important role is played by the so-called \textit{Planck length}, the unique
 combination, with dimensions of length, of Newton's gravitational constant ($G$), Planck's constant divided by $2\pi$ ($\hbar$),
and the speed of light ($c$). It is given by $\ell_{\Pl}:= \sqrt{\frac{G\hbar}{c^3}}$, which is approximately
$1.616 \times 10^{-35} m$, or roughly ten sextillionths
of (or $10^{-20}$ times) the diameter of a proton.
Given a surface $S$, in terms of the Planck length, its area as determined by a spin network state with the labels
$\gamma, \{j_\link, n_{\node\link}\}$ is
\begin{equation}
\label{areaeq}
A(S) = \sum_{\link \text{ intersecting } S} 8 \pi \ell_{\Pl}^2 \BI \sqrt{j_\link(j_\link + 1)}.
\end{equation}
where $\BI$ is a certain positive real number referred to as the Barbero-Immirzi parameter \cite{barbero1995, immirzi1997, meissner2004, abbdv2009}.
(See figure \ref{qarea}.)
Given a three-dimensional region $R$ in space, its volume is
%
%
%
\begin{equation}
\label{voleq}
V(R) = \frac{(8\pi \BI)^{3/2} \ell_{Pl}^3}{4\sqrt{3}}
\sum_{\substack{\node \text{ nodes of } \gamma \\ \text{ in } R}} \quad
\sqrt{\Bigg|\sum_{\substack{\link, \link', \link''\\ \text{at }\node}} j_{\link} j_{\link'} j_{\link''} \,
n_{\node\link}\cdot (n_{\node\link'} \times n_{\node\link''})\Bigg|}
\end{equation}
where the sum over $\link, \link', \link'$ is over all triples of
links in $\gamma$ starting or ending at the node $\node$.
%
%

\subsection{Interpretation of spin networks in terms of the dual complex}
\label{sndual_sect}

\begin{figure}[t]
\begin{minipage}[t]{2.85in}
\centering
\setlength{\unitlength}{1in}
\begin{picture}(1.5,1.5)
\includegraphics[height = 1.5in]{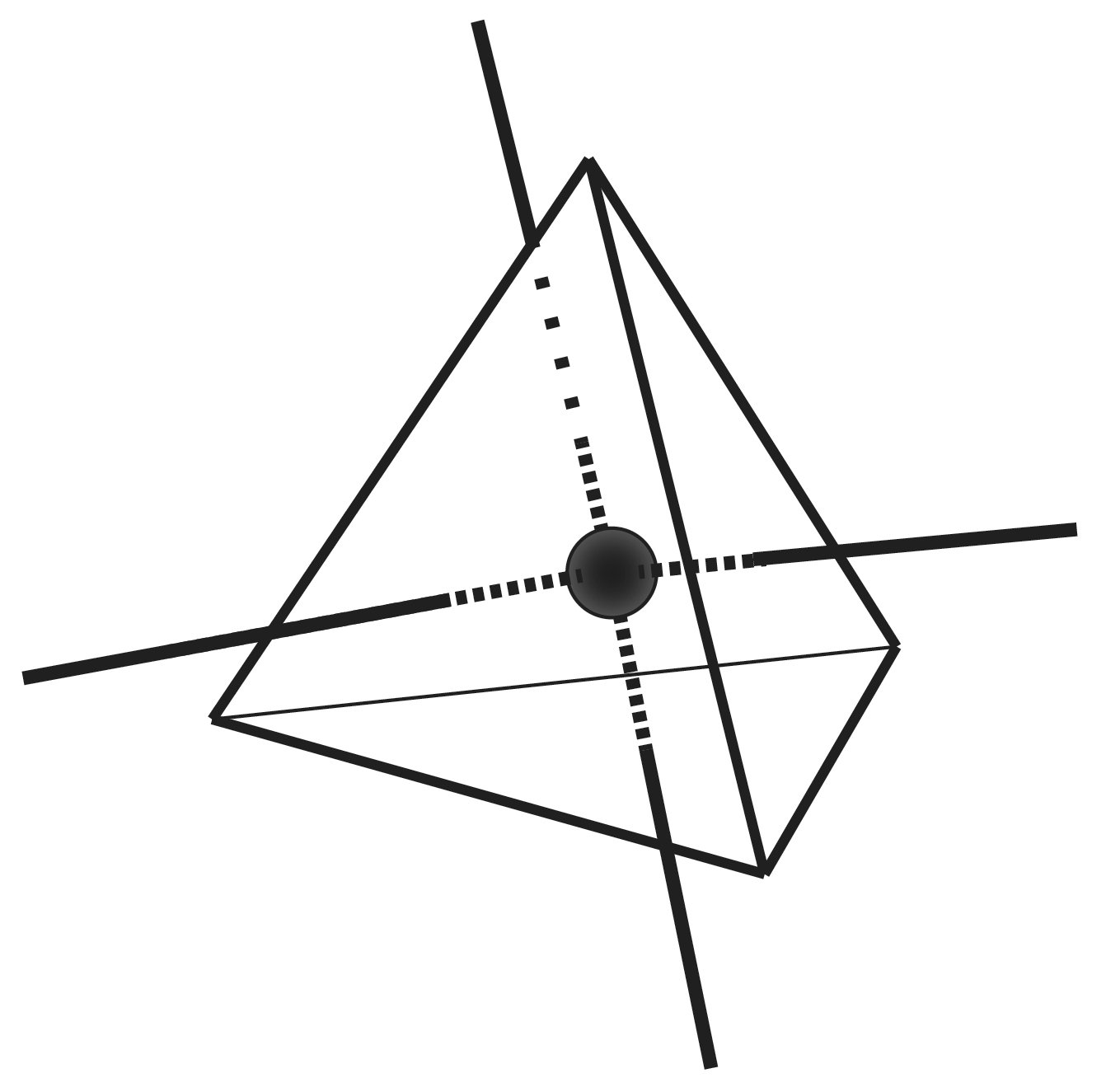}
\put(-1.1,0.9){$R$}
\put(-0.81,1.35){$\gamma$}
\put(-0.83,0.75){$\node$}
\end{picture}
\caption{\label{qvolume}A three-dimensional region $R=\node\dual$ is said to be \textit{dual} to a node $\node$ of $\gamma$
if it contains $\node$ but no other node of $\gamma$.}
\end{minipage}
\quad
\begin{minipage}[t]{2.85in}
\centering
\setlength{\unitlength}{1in}
\begin{picture}(1.5,1.7)
\includegraphics[height=1.5in]{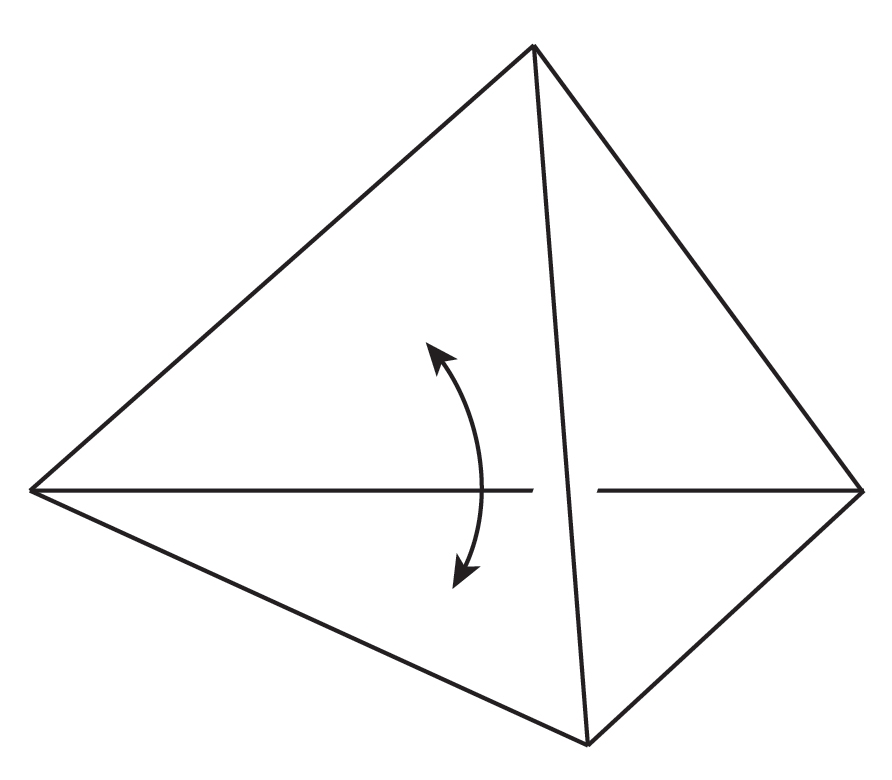}
\put(-1.25,1.05){$\node\dual$}
\put(-0.9,0.9){$\link\dual$}
\put(-0.78,0.28){$\tilde{\link}\dual$}
\put(-0.75,0.6){$\theta$}
\end{picture}
\caption{\label{interioranglefig}
The interior angle $\theta$ between the two faces $\link\dual$ and $\tilde{\link}\dual$ of the 3-cell $\node\dual$,
as determined by the LQG spin-network labels, is given by equation (\ref{thetaeq}).
}
\end{minipage}
\end{figure}

The extraction of information about geometry from the quantum labels
$j_\link$, $n_{\node\link}$ can be
systematized using what is called a \textit{dual cell complex}.  For each link $\link$, a surface (a two dimensional region) $S=\link\dual$ is said to be \textit{dual} to $\link$ if it intersects $\link$ at one point, but intersects no other link of $\gamma$.
For each node $\node$, a three-dimensional region $R=\node\dual$ is said to be \textit{dual} to $\node$ if it contains 
$\node$ but no other node of $\gamma$ (see figure \ref{qvolume}).
If one chooses such a dual for each link and node in the graph $\gamma$,  and if these are chosen such that they all ``fit together'' --- that is, such that the boundary of each chosen three-dimensional region $\node\dual$ consists entirely of chosen two-dimensional region $\link\dual$, then the set of all the chosen regions $\node\dual$, $\link\dual$
form a \textit{cell-complex} which is said to be \textit{dual to} $\gamma$, and
which we denote by $\gamma\dual$.  In this case, we refer to $\node\dual$ and
$\link\dual$ as \textit{cells} of $\gamma\dual$; more specifically one uses the terms
\textit{3-cell} and \textit{2-cell}, respectively, according to the dimension of the region.
From (\ref{areaeq}) the spin $j_\link$ on a link $\link$ determines the area of the surface $\link\dual$ 
dual to it by the formula
\begin{equation}
\label{linkarea}
A(\link\dual) = 8 \pi \ell_{\Pl}^2 \BI \sqrt{j_\link(j_\link + 1)}.
\end{equation}
From (\ref{voleq})
the quantum labels $n_{\node\link}$ at a given node $\node$ determine the volume of the region $\node\dual$ 
dual to it via the formula
\begin{displaymath}
V(\node\dual) = \frac{(8\pi \BI)^{3/2} \ell_{Pl}^3}{4\sqrt{3}}
\sqrt{\Bigg|\sum_{\substack{\link, \link', \link''\\ \text{at }\node}} j_{\link} j_{\link'} j_{\link''} \,
n_{\node\link}\cdot (n_{\node\link'} \times n_{\node\link''})\Bigg|}.
\end{displaymath}
In addition to this, given a node $\node$ and two links $\link,
\tilde{\link}$ incident at it, one can ask what is the \textit{angle} $\theta =\theta[\node\dual, \link\dual, \tilde{\link}\dual]$ between the dual surfaces $\link\dual, \link'\dual$ within the dual
region $\node\dual$.  In fact, it is given by the formula
\begin{equation}
\label{thetaeq}
\cos \left(\theta[\node\dual, \link\dual, \tilde{\link}\dual]\right) = - n_{\node\link} \cdot n_{\node\tilde{\link}} .
\end{equation}
(See figure \ref{interioranglefig}.)
These areas, volumes, and interior angles form the basic quantities from which the quantum geometry is
constructed.  We will go into more detail about this in section \ref{qstgeom_sect}.

There is of course a great deal of choice in the complex $\gamma\dual$ dual to $\gamma$.  However,
given $\gamma$, the \textit{connectivity} of the parts of $\gamma\dual$ is uniquely determined --- that is, which lower dimensional cells are on the boundary of each higher dimensional cell \textit{is} uniquely determined.
If $\node$ is on the boundary of $\link$
(meaning, in this case, an endpoint of $\link$), then $\link\dual$ is on the boundary of $\node\dual$.
Another way of saying this, in mathematical terms, is that the \textit{topology} of $\gamma\dual$ is unique,
and it is in this sense that we can speak unambiguously of ``the complex $\gamma\dual$ dual to $\gamma$.''

\begin{figure}
\begin{center}
\setlength{\unitlength}{1in}
\begin{picture}(2,2)
\includegraphics[height=2in]{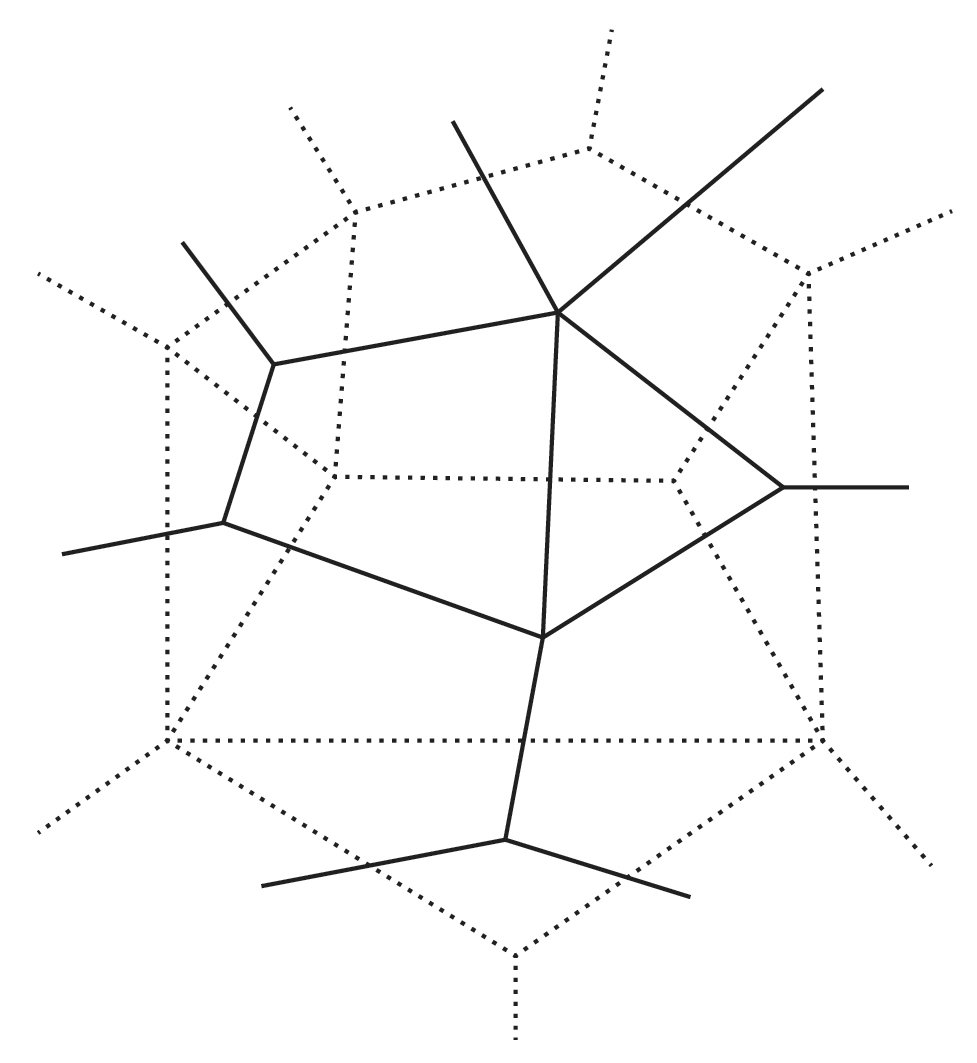}
\end{picture}
\end{center}
\caption{\label{dualcomplex_fig}
Example of dual cell complexes in two dimensions.
The solid line complex and the dotted line complex are dual to each other.
}
\end{figure}

We have here discussed dual cells and dual cell complexes in three dimensions.
However, these ideas can be formulated in any dimension.  If one is working in
an $N$-dimensional space, and one has an $M$-dimensional surface $S$, a surface $S\dual$
is said to be \textit{dual} to $S$ if it has dimension $N-M$ and intersects $S$ at exactly
one point.  See table \ref{cellduals} later on in the chapter for examples of dual surfaces
in 2 and 4 dimensions.  In the case of 2 dimensions, one can visualize the idea
of a dual cell complex with more completeness and ease.
For the purpose of illustration, we include in figure \ref{dualcomplex_fig} an example of a dual complex in two dimensions.
%
%

\subsection{Histories of spin networks: Spin foams}
\label{lqgsf_sect}

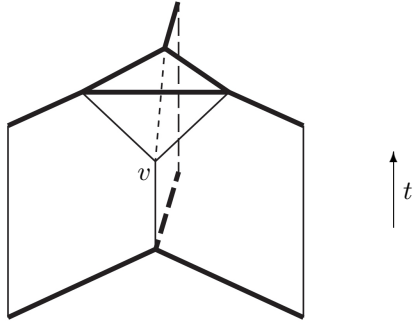
\begin{figure}
\begin{center}
\setlength{\unitlength}{1in}
\begin{picture}(1.7,1.7)
\includegraphics[height=1.7in]{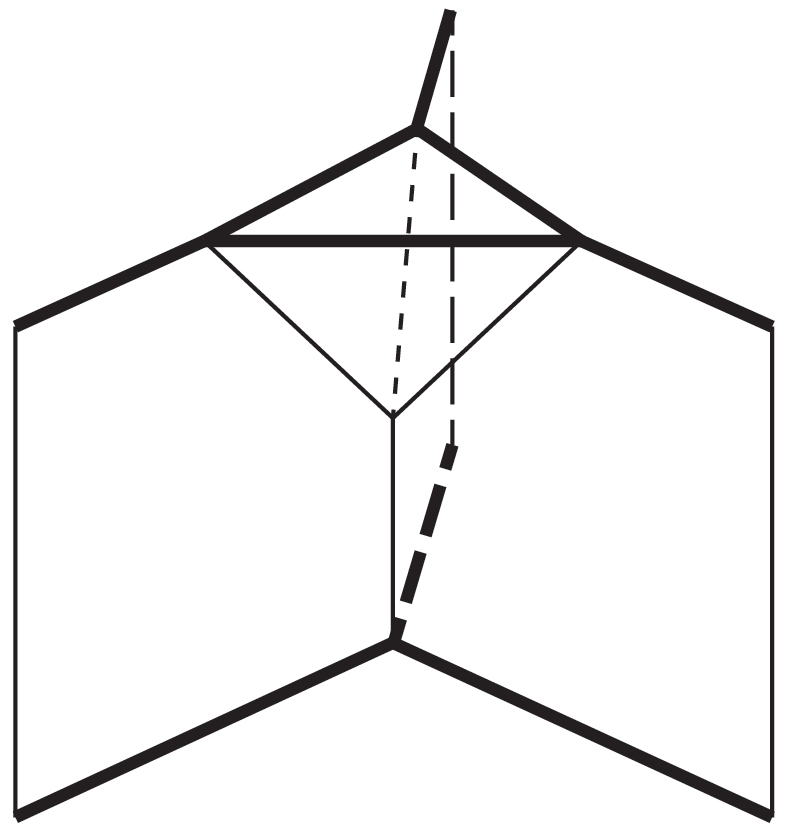}
\put(-0.9,0.75){$v$}
\put(0.44,0.5){\vector(0,1){0.4}}
\put(0.49,0.65){$t$}
\end{picture}
\end{center}
\caption{\label{vertexfig}
A single node splits into three nodes, creating a spin foam vertex.
}
\end{figure}

\begin{figure}
\begin{center}
\setlength{\unitlength}{1in}
\begin{picture}(1.5,2.5)
\includegraphics[height=2.5in]{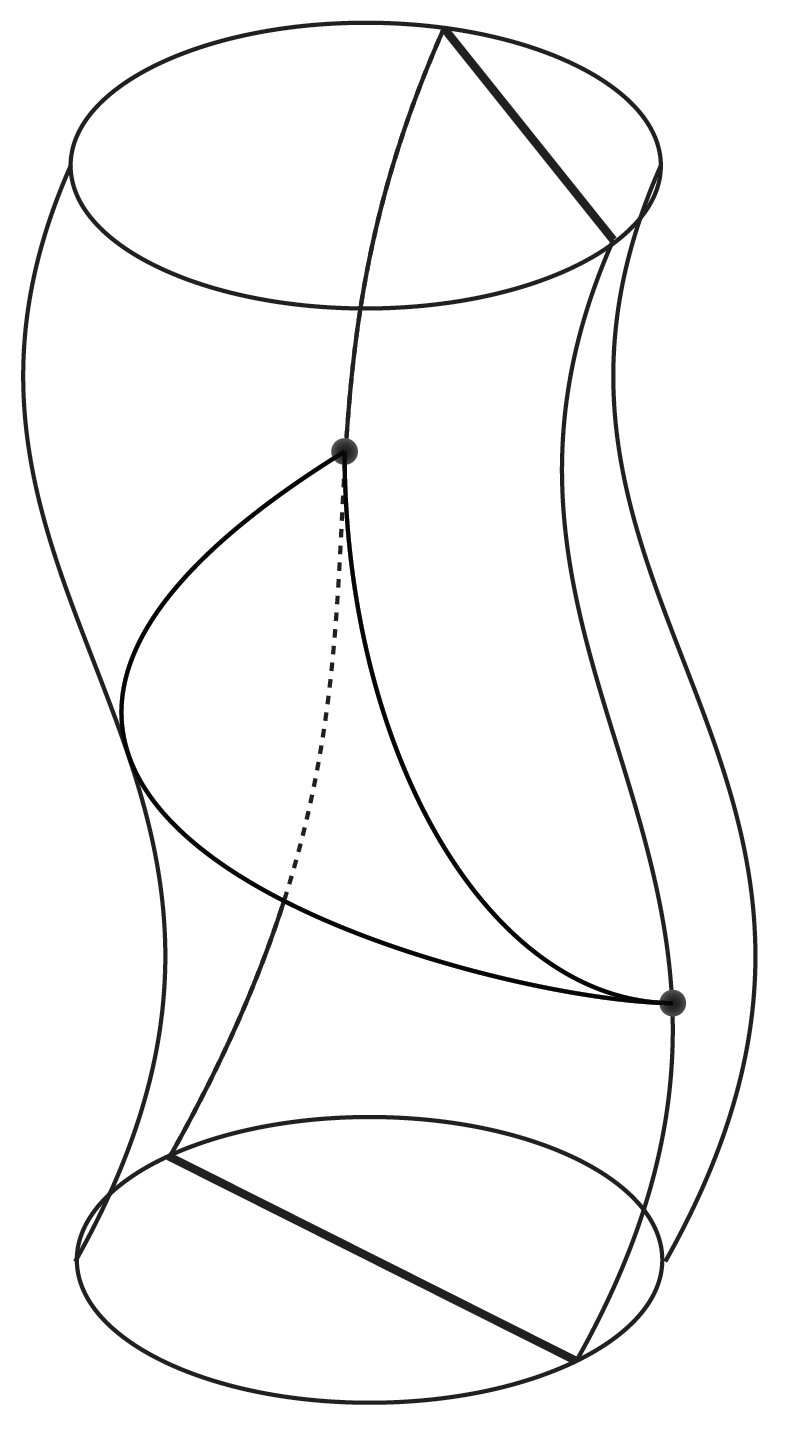}
%
%
\put(-1.25,0.65){$f_1$}
\put(-0.94,0.68){$e_1$}
\put(-0.67,0.63){$f_2$}
\put(-0.328,0.56){$e_2$}
\put(-0.17,0.72){$v_1$}
\put(-0.05,1.1){$f_3$}
\put(-1,1.2){$f_4$}
\put(-0.58,1.05){$e_4$}
\put(-0.28,1.2){$e_5$}
\put(-1.05,1.425){$e_3$}
\put(-0.62,1.47){$f_5$}
\put(-1.47,1.7){$f_6$}
\put(-0.92,1.73){$v_2$}
\put(-0.68,2.15){$e_6$}
\end{picture}
\end{center}
\caption{\label{lqgsf_fig}
A spin foam is a history of a spin network.  It forms a \textit{two-complex}, with the links of the spin network
sweeping out faces, and the nodes of the spin network sweeping out edges.  Each face $f$  in the spin foam inherits the
spin on the corresponding link, and each edge $e$ in the spin foam inherits the set of unit three dimensional vectors labeling
the corresponding node.  The face spins are now denoted $j_f$, and the three dimensional vectors are now denoted
$n_{ef}$
}
\end{figure}
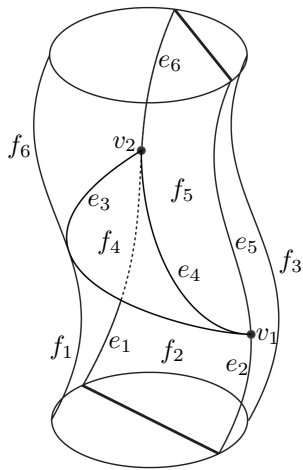

Histories of three-dimensional spin networks  $|\gamma, \{j_\link, n_{\node\link}\} \rangle$,
become four dimensional objects.  The one-dimensional links of the graphs $\gamma$ become
two-dimensional `faces' $f$, and the zero dimensional nodes of the graphs become one-dimensional `edges' $e$.
Places in the history where a node splits into multiple nodes, or multiple nodes combine are called
\textit{vertices}. (See figure \ref{vertexfig}.)
The set of all such faces, edges, and vertices of a given history together form the \textit{spin foam two-complex} of the history,
which we usually
denote $\mathcal{F}$. (See figure \ref{lqgsf_fig}.)
Each face $f$ inherits the half integer spin $j_{f}$ labeling the  link of which it is the history, and
each edge $e$ inherits the set of unit vectors associated to the node of which it is a history, one unit
vector $n_{ef}$ for each edge $e$ and face $f$ incident at $e$.
The spin foam two-complex $\mathcal{F}$, together with these labels, is referred to as a \textit{spin foam}.
Specifically, with this choice of labels, we will call it a \textit{loop quantum gravity spin foam}.
Each such spin foam represents, in a precise sense to be reviewed in section \ref{qstgeom_sect},
a \textit{quantum space-time geometry}.

\subsection{Spin foam amplitudes}

In order to specify the quantum dynamics, a \textit{probability amplitude}
must be specified for each spin foam --- that is, a probability amplitude for each history
of quantum gravity states, each `quantum space-time'.
This amplitude should be, in an appropriate semiclassical limit, equal to (a possible real coefficient times) the usual
Feynman prescription of the exponential of $i$ times the classical action, as  reviewed in
section \ref{qhist_sect}.

It turns out, from experience with simple theories in four space-time dimensions and gravity in three space-time dimensions \cite{os1991, ooguri1992},
one expects this amplitude to be of the form
\begin{equation}
\label{sfamp}
\hamp(\mathcal{F}, \{j_f, n_{ef}\})  =
\left(\prod_{f \in \mathcal{F}} \hamp_f\right)
\left(\prod_{e \in \mathcal{F}} \hamp_e\right)
\left(\prod_{v \in \mathcal{F}} \hamp_v \right)
\end{equation}
where for each face $f$, edge $e$, and vertex $v$,
$\hamp_f$, $\hamp_e$, and $\hamp_v$ are referred to as the \textit{face}, \textit{edge}, and \textit{vertex} amplitudes,
respectively.  This form of the probability amplitude is called the \textit{spin foam Ansatz}.
Here, $\hamp_f$ is a function of the spin $j_f$ alone, $\hamp_e$ is a function of the quantum labels associated
to the edge $e$ as well as to the faces incident at $e$, and $\hamp_v$ is a function of the quantum labels associated
to the edges and faces incident at the vertex $v$. From experience with the above mentioned
simple models, $\hamp_f$ and $\hamp_e$ are expected to be real, and $\hamp_v$ complex.  Thus, one
expects the exponential of $i$ times the action to arise almost entirely from the vertex amplitudes
alone. It is for this reason that the vertex amplitude is usually considered the most important one.
Furthermore, the vertices are where the spin network `changes' in the history,  and hence where
`interesting dynamics' is taking place. Thus, in a sense, it is not surprising that the vertex amplitude usually
turns out to be the most important factor in the probability amplitude.

\section{Deriving the amplitude via a simpler theory}

How should one determine the different factors $\hamp_f$, $\hamp_e$, and $\hamp_v$ appearing in the
probability amplitude (\ref{sfamp})? The strategy used by the spin foam community is a bit
indirect: We first construct the spin foam amplitude for a very simple toy theory, called BF theory.
The spin foam dynamics of BF theory is very well understood.  One then uses the fact that
general relativity can be obtained from BF theory by imposing extra constraints, called
\textit{simplicity constraints}, an idea which traces back to the work of Plebanski \cite{plebanski1977}.
The non-trivial task in constructing a spin foam model
is then reduced to the question of how these simplicity constraints should be imposed in the quantum theory.

We begin this section by reviewing a minimum necessary to understand what is BF theory, what are the
simplicity constraints, and how Einstein's theory of gravity can be recovered from these.
We then review the quantum mechanics of BF theory, and then discuss the version of the
quantum simplicity constraints now predominant in the literature.  (For the first method of
imposing quantum simplicity which was previously predominant, and which layed the foundations for the modern method,
see \cite{bc1997, bc1999}.)

\subsection{BF theory and gravity}
\label{BFgravity_sect}

BF theory is a theory with a maximal number of gauge symmetries.
Recall that a \textit{gauge symmetry} is a transformation that does not change the physical state of the system,
but only changes the variables used to describe it.   That is, the presence of gauge symmetries
in a theory indicates a redundancy in the variables used to describe the system. The simplest example
of a gauge symmetry is a transformation of the vector potential of the magnetic field:
Given a vector potential $\vec{A}$ and a function $\chi$ on space, the new vector potential
$\vec{\tilde{A}}:= \vec{A} + \nabla \chi$
determines exactly the same magnetic field, and hence the same physical state of the system.
In the case of general relativity, the gauge transformations are \textit{space-time coordinate transformations},
reflecting the physical fact that space-time coordinates have no intrinsic meaning in the theory: Space-time coordinates
are only tools of convenience, used to aide in describing  physical fields.  The more gauge symmetries one has in a system,
the less the variables of the theory contain real, physical information.  BF theory has so many gauge symmetries that in fact
the variables of the theory contain \textit{no} local information.  This is why BF theory is so simple and why the
corresponding spin foam quantum theory is so well understood.
%
%
Such simple theories like BF theory which have no local
physical degrees of freedom are referred to as \textit{topological field theories}.

To introduce the basic variables of BF theory, we first recall a notation usually used for matrices:
Given a matrix $M$, one denotes its element in the $i$th row and $j$th column by $M_{ij}$.
When one allows more than just two indices, one obtains a generalization of matrices which we shall call
\textit{arrays}.  The basic variables of BF theory are two fields of arrays on space-time,
denoted $\Sigma_{\mu\nu}^{IJ}(x)$ and $\omega_{\mu}^{IJ}(x)$,
where the indices $\mu,\nu,I,J$ take the values $0,1,2,3$.  (There are more specific terms for these
types of fields of arrays, which indicate certain transformation properties, but
we have chosen to avoid these terms, because we wish to avoid talking about transformation properties
which are not necessary for the discussion in this chapter.)
%
%
In terms of these variables, the action for BF theory is given by
\begin{eqnarray}
\label{BFaction}
S_{BF} &=& \frac{1}{32\pi G} \int \epsilon^{\mu\nu\sigma\rho}
\left(\half \epsilon_{IJKL} \Sigma_{\mu\nu}^{KL} + \frac{1}{\BI} \eta_{IK} \eta_{JL} \Sigma_{\mu\nu}^{KL} \right)
 F_{\sigma\rho}^{IJ}  d^4 x\\
\nonumber
&=:& \frac{1}{2}\int \epsilon^{\mu\nu\sigma\rho} B_{\mu\nu IJ} F_{\sigma\rho}^{IJ} d^4 x
\end{eqnarray}
%
%
where $\epsilon_{IJKL}$, $\epsilon^{\mu\nu\sigma\rho}$ both denote the `Levi-Civita array', defined uniquely by the properties
$\epsilon_{0123} = 1$ and that when any two indices of $IJKL$ (respectively $\mu\nu\sigma\rho$) 
are interchanged, $\epsilon_{IJKL}$ (respectively $\epsilon^{\mu\nu\sigma\rho}$) changes
by minus sign --- for example, $\epsilon_{IJKL} = - \epsilon_{JIKL}$.
Furthermore, here and throughout the rest of this section we use the Einstein summation convention:
when a given index appears twice in an expression, once up and once down, summation shall be implied over all
possible values of the given index.  In the final expression above, we have defined $B_{\mu\nu}^{IJ}$ to
be the quantity in parentheses, and $F_{\sigma\rho}^{IJ}$ denotes the \textit{field strength} (or \textit{curvature}) of
$\omega_\mu^{IJ}$, defined by
\begin{displaymath}
F_{\sigma\rho}^{IJ} :=
\frac{\partial \omega_\rho^{IJ}}{\partial x^\sigma} - \frac{\partial \omega_\sigma^{IJ}}{\partial x^\rho}
+ \eta_{KL} \left( \omega_\sigma^{IK} \omega_\rho^{LJ}
- \omega_\rho^{IK} \omega_\sigma^{LJ}\right).
\end{displaymath}
(Note that in the spin foam literature, sometimes $B_{\mu\nu}^{IJ}$ is defined to be only the first term of the expression in
parentheses in (\ref{BFaction}).)

The simplicity constraint, in its simplest and most important sense, is just the requirement that
there exist a matrix field $e^I_\mu(x)$ such that
\begin{equation}
\label{simp}
\Sigma_{\mu\nu}^{IJ}(x) = \pm \left( e^I_\mu(x) e^J_\nu(x) - e^J_\mu(x) e^I_\nu(x) \right)
\end{equation}
at each  space-time point $x$.
The simplicity constraint is thus a constraint on the field $\Sigma_{\mu\nu}^{IJ}(x)$;
when $\Sigma_{\mu\nu}^{IJ}(x)$ satisfies this constraint, one says $\Sigma_{\mu\nu}^{IJ}(x)$ is `simple'.
When $\Sigma_{\mu\nu}^{IJ}(x)$ is simple, the fields describing the system are just $e^I_\mu$ and $\omega_\mu^{IJ}$.
What is important to know is that these fields in fact just describe a geometry for space-time.
In terms of the metric tensor $g_{\mu\nu}$ used in much of this handbook,
%
%
this geometry is just $g_{\mu\nu} = \eta_{IJ }e^I_\mu e^J_\nu$ where $\eta_{IJ}$ is the diagonal matrix
with $\eta_{11} = \eta_{22} = \eta_{33} = 1$, and $\eta_{00} = \pm1$ depending on whether one is considering
Euclidean ($+1$) or Lorentzian ($-1$) gravity.  
$e^I_\mu$ is referred to as a \textit{co-tetrad}.
The terms ``Euclidean'' and ``Lorentzian'' gravity are a bit misleading.
In fact, there is only one Einstein theory of gravity describing the real world, and that is what we are calling here
``Lorentzian'' gravity.  ``Euclidean gravity'' is a simplified model very closely related to, but in certain ways simpler than,
Lorentzian gravity.  It is often used for ``practice'' when investigating quantum gravity.  Essentially, in Euclidean gravity
one treats time as though it were just a fourth dimension of space.
In this chapter, we consider spin foam quantizations
of both of these models of gravity.

\subsection{Spin foams of BF theory}
\label{sfbf_sect}

We have already described the spin foams arising from loop quantum gravity.  We next describe
spin foams for BF theory. These again arise as histories of labels of
corresponding canonical quantum states,
just as the loop quantum gravity spin foams of section \ref{lqgsf_sect} arose as histories of labels of the
canonical Livine-Speziale spin network states of loop quantum gravity.
As mentioned, there are two basic variables of BF theory, $\Sigma_{\mu\nu}^{IJ}$ and $\omega_\mu^{IJ}$.
The restrictions of these fields to a given instant-time hypersurface are canonically conjugate, so that,
to have a complete set of canonical states, it is sufficient to consider states peaked on
$\Sigma_{\mu\nu}^{IJ}$ \textit{or} $\omega_\mu^{IJ}$, but not both.
The particular choice of canonical states we use for defining BF spin foams
are peaked on the variable $\Sigma_{\mu\nu}^{IJ}$ and
are closely related to the Livine-Speziale spin networks \cite{engle2011, bdfgh2009, bdfhp2009}.
This choice will facilitate imposing the simplicity constraint, as well as be important
for taking the semiclassical limit of the theory.

Exactly as in the case of the loop quantum gravity spin foams, each BF spin foam is first labeled
by a \textit{spin foam two-complex}, with faces, edges, and vertices (as in figure \ref{lqgsf_fig}).
However, now the labels on the faces and edges are different.
As mentioned at the end of the last section, in quantum gravity, often one considers first the simplified
theory of Euclidean gravity for practice, before considering the actual Lorentzian gravity corresponding
to reality.  Spin foams is no exception.  The spin foam quantum labels for BF theory are different depending
on whether one considers Euclidean or Lorentzian gravity.  In the Euclidean case, each face $f$ is labeled by
two half-integers $j_f^+, j_f^- = 0,\frac{1}{2}, 1, \frac{3}{2}, \dots$, and for each edge $e$ in the boundary of $f$,
one has a further half integer $j_{ef}$ and a unit three dimensional vector $n_{ef}$.
In the Lorentzian case, each face $f$ is labeled by a real number $p_f$ and a half-integer $k_f$,
and for each edge $e$ in the boundary of $f$ one again has a half integer $j_{ef}$ and a unit
three dimensional vector $n_{ef}$.  See table \ref{BFSFlabels}.
\begin{table}[t]
\begin{centering}
\begin{tabular}{lcc}
&for each face $f$ & for each edge $e \in \partial f$ \\
\cline{2-3}
Euclidean & $j_f^+, j_f^-$ &  $j_{ef}, n_{ef}$  \\
\cline{2-3}
Lorentzian &  $p_f, k_f$ & $j_{ef}, n_{ef}$ \\
\cline{2-3}
\end{tabular}
\caption{
Labels on each face and on each edge bounding each face, for the BF spin foam model used
in the case of Euclidean and Lorentzian gravity, respectively.
}
\label{BFSFlabels}
\end{centering}
\end{table}
For convenience we let $\mathcal{L}$ denote the appropriate set of possible labels on the BF spin foam:
$\{j^+_f, j^-_f, j_{ef}, n_{ef}\}$ in the Euclidean case,
and $\{p_f, k_f, j_{ef}, n_{ef}\}$ in the Lorentzian case.

In terms of these labels, the amplitude for a single BF spin foam decomposes as in the spin foam Ansatz (\ref{sfamp}),
with certain expressions for the corresponding face, edge, and vertex amplitudes $\hamp_f^{BF}$,
$\hamp_e^{BF}$, $\hamp_v^{BF}$.
What is important for this chapter is that the vertex amplitude can be expressed in terms of integrals over certain
\textit{groups}. Let $\mathcal{G}$ denote the space of all $4 \times 4$ matrices $G^I{}_J$ such that
\begin{displaymath}
G^K{}_I G^L{}_J \eta_{KL} = \eta_{IJ}.
\end{displaymath}
where $\eta_{IJ}$ is again the diagonal four by four matrix with diagonal components $\eta_{11}=\eta_{22}=\eta_{33}=1$
and $\eta_{00} = +1$ or $-1$ depending on whether one is considering Euclidean or Lorentzian gravity.
For the case of Lorentzian gravity, a matrix $G$ satisfies the above equation if and only if
its action on a given set of four space-time coordinates is a \textit{Lorentz transformations};  in this case
$\mathcal{G}$ is called the \textit{Lorentz group}.  In the case of Euclidean
gravity, $\mathcal{G}$ is the group of \textit{four dimensional Euclidean rotations}.
(In fact, the group $\mathcal{G}$ is directly related to the labels on the faces of the BF spin-foam:
Each pair $(p_f, k_f)$ labels a unitary irreducible representation of the Lorentz group, and each pair
$(j_f^+, j_f^-)$ labels a unitary irreducible representation of the group of four dimensional Euclidean rotations.
This is similar to the way the angular momentum quantum number $j$ in basic quantum mechanics
labels irreducible representations of the spatial rotation group.)
There is a way to define integrals over the group of matrices $\mathcal{G}$.
The vertex amplitude $\hamp_v^{BF}$ of BF theory
can be expressed in terms of nested integrals over such matrices, one such
integral for each edge $e$ incident at the given vertex $v$:
\begin{equation}
\label{integform}
\hamp_v^{BF}(\mathcal{L}) = \left(\prod_{e\text{ incident at }v} \int_\mathcal{G} dG_{ve}\right)
\tilde{\hamp}_v^{BF}(\mathcal{L}, \{G_{ve}\}).
\end{equation}
One can think of the spin foam two-complex $\mathcal{F}$, together with the labels $\mathcal{L}$ \textit{and}
the group matrices $\{G_{ve}\}$ as labelling a sort of `augmented' history,
and $\tilde{\hamp}_v^{BF}(\mathcal{L}, \{G_{ve}\})$ is the probability amplitude associated to this history.
It is these augmented histories that will have a complete interpretation in terms of the classical
variables of BF theory, as we will see in the next subsection. Beyond the above general form (\ref{integform}), the details
of the vertex amplitude will not be needed in this chapter.

\subsection{Dual cell complex}
\label{dual_sect}

To interpret the quantum labels for the BF spin foams in terms of classical BF theory, we use the same strategy as that used in
section \ref{sndual_sect} to interpret spin-network labels:  We again use of the notion of a \textit{dual cell complex},
except now in one dimension higher.
In this subsection we explicitly spell out this duality in the four dimensional case,
lifting the duality presented in section \ref{sndual_sect} from space to space-time.

Recall that each spin foam is first of all labeled by a spin foam \textit{two-complex} $\mathcal{F}$, consisting in vertices,
edges, and faces, which fit together. For each vertex $v$ in $\mathcal{F}$,
a four dimensional region $v\dual$ is said to be \textit{dual} to $v$ if it contains $v$ and no other vertices of $\mathcal{F}$.
For each edge $e$ in $\mathcal{F}$, a three-dimensional hypersurface $e\dual$ is said to be \textit{dual} to $e$ if it
intersects $e$ in exactly one point, and intersects no other edges in $\mathcal{F}$.  For each face $f$ in $\mathcal{F}$,
a two-dimensional surface $f\dual$ is said to be \textit{dual} to $f$ if it intersects $f$ at one point, and intersects no other
faces in $\mathcal{F}$. (For comparison with examples of dual cells when working in lower dimensions, see table \ref{cellduals}.)
\newcommand{\myarrow}{\begin{minipage}{0.3in}$\leftrightarrow$\\
\vspace{0.3cm}\dummy\end{minipage}}
\begin{table}[t]
\begin{centering}
\begin{tabular}{c@{\hspace{1.5cm}}c@{\hspace{1.5cm}}c}
\underline{2-D} & \underline{3-D} & \underline{4-D} \\ \\
\includegraphics[height=0.4in]{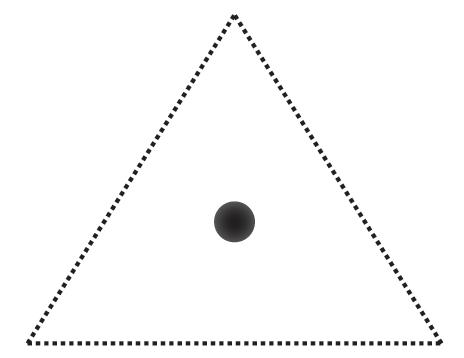} &
\includegraphics[height=0.4in]{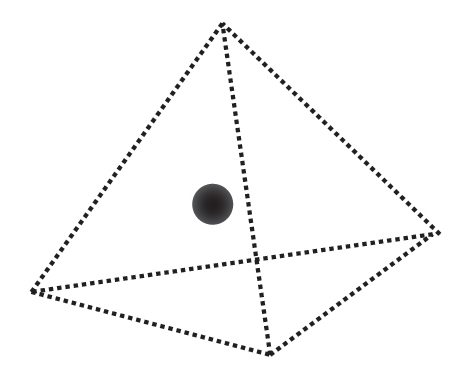} &
\includegraphics[height=0.4in]{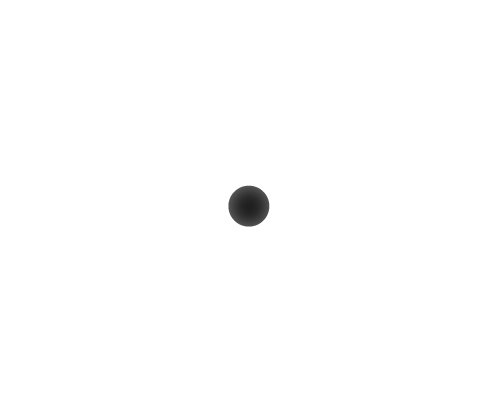}
\myarrow
\includegraphics[height=0.4in]{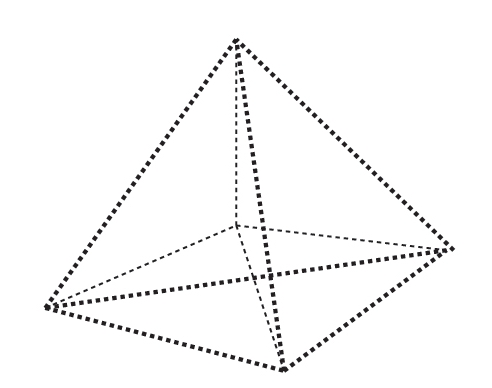}\\
0+2=2 & 0+3=3 & 0+4 = 4 \\ \\
\includegraphics[height=0.4in]{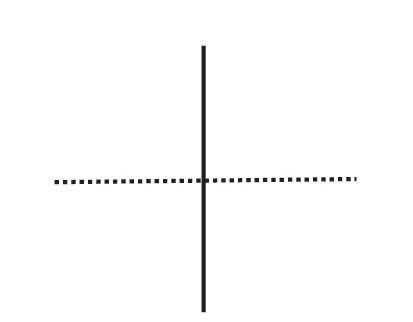} &
\includegraphics[height=0.4in]{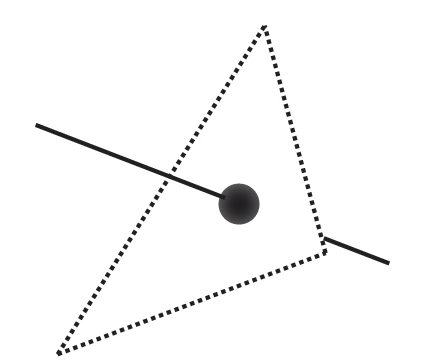} &
\includegraphics[height=0.4in]{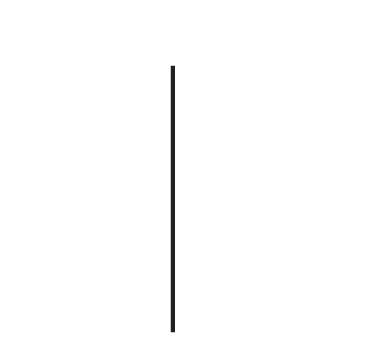}
\myarrow
\includegraphics[height=0.4in]{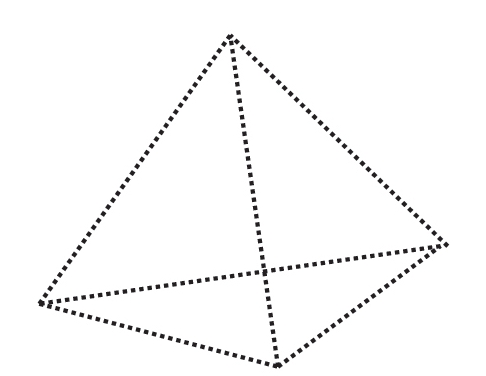}\\
1+1=2 & 2+1=3 & 1+3=4 \\ \\
\includegraphics[height=0.4in]{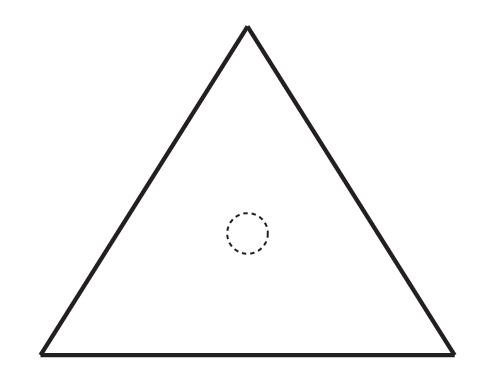} &  &
\includegraphics[height=0.4in]{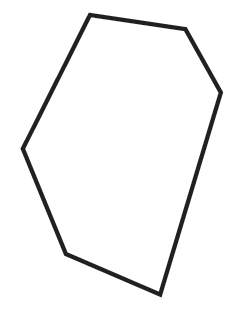}
\hspace{0.1cm} \myarrow \hspace{-0.3cm}
\includegraphics[height=0.3in]{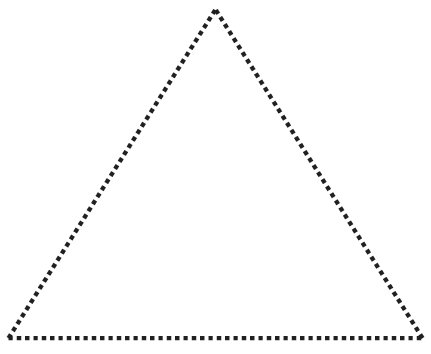}\\
2+0 = 2 & & 2+2 = 4
\end{tabular}
\caption{
Examples of dual cells in two, three, and four dimensions.  For each pair of dual cells, the dimensionality of the cells
add up to the total dimensionality of the ambient space, and intersect in one point.  In the four dimensional case, the fact that
dual cells intersect in one point cannot be depicted.
}
\label{cellduals}
\end{centering}
\end{table}
If one choses such a dual for each vertex, edge, and face in $\mathcal{F}$, and if these are chosen such that
they all ``fit together'' --- that is, such that
the boundary of each chosen four-dimensional region $v\dual$ consists entirely of chosen three-dimensional
hypersurfaces $e\dual$, and the boundary of each three dimensional hypersurface $e\dual$ consists entirely of chosen two-dimensional surfaces $f\dual$, then the set of all the chosen regions $v\dual$, $e\dual$, $f\dual$ form a \textit{cell complex} which is said to be
\textit{dual to} $\mathcal{F}$, and which we denote by $\mathcal{F}\dual$.
In this case we refer to $v\dual$, $e\dual$, and $f\dual$ as \textit{cells} of $\mathcal{F}\dual$;
more specifically one uses the terms \textit{4-cell}, \textit{3-cell}, and \textit{2-cell}, respectively, according
to the dimension of the region.
Once again, though there is a great deal of choice in such a complex $\mathcal{F}\dual$ dual to $\mathcal{F}$, the
\textit{connectivity} of the parts of $\mathcal{F}\dual$ is uniquely determined --- that is, the \textit{topology} of $\mathcal{F}\dual$ is unique.

\subsection{Interpretation of the labels}
\label{BFinterp}

To interpret these labels, we arbitrarily fix a coordinate system $x^\mu$ in each 4-cell $v\dual$ such
that, in this coordinate system, each 3-cell $e\dual$ and 2-cell $f\dual$ bounding $v\dual$ is \textit{planar}.
%
%
No physical quantities arising from the constructions that follow depend on this choice of coordinates in each $v\dual$.
The classical field $\Sigma_{\mu\nu}^{IJ}$ corresponding to a given `augmented' BF spin foam $(\mathcal{F}, \mathcal{L}, G_{ve})$
is then constant in each 4-cell $v\dual$ (in the coordinates $x^\mu$ fixed in each $v\dual$).
Let $(\Sigma_v)_{\mu\nu}^{IJ}$ denote the constant value taken by $\Sigma_{\mu\nu}^{IJ}(x)$ in the cell $v\dual$.
The labels $\{j_{ef}, n_{ef}, G_{ve}\}$ are then related to $(\Sigma_v)_{\mu\nu}^{IJ}$ by
\begin{equation}
\label{leq}
8\pi \ell_\Pl^2 j_{ef} n_{ef}^i
= \left((G_{ve})^0{}_L (G_{ve})^i{}_M + \frac{s}{\BI} (G_{ve})^j{}_L (G_{ve})^k{}_M\right)
\int_f \Sigma_v^{LM}
\end{equation}
for $(i,j,k) = (1,2,3), (3,1,2),(2,3,1)$, where $s =+1$ for Euclidean gravity and $-1$ for Lorentzian gravity, and where, recall,
$\ell_\Pl$ denotes the Planck length.
The remaining labels, $\{j^+_{f}, j^-_{f} \}$ in the Euclidean case
and $\{p_f, k_f\}$ in the Lorentzian case, are related to the classical field $\Sigma_{\mu\nu}^{IJ}(x)$ by
%
%
\begin{eqnarray}
\label{caseq}
\eta_{IK} \eta_{JL} \left(\int_f \Sigma^{IJ}\right) \left( \int_f \Sigma^{KL}\right)
\hspace{-0.2cm}&=& \hspace{-0.2cm} \left\{ \begin{array}{ll} C_E \left(\frac{(j^+_f)^2}{(\BI + 1)^2}
+ \frac{(j^-_f)^2}{(\BI - 1)^2}\right) & \text{in Eucl. case}\\
C_L
\left(k_f^2 - p_f^2 + \frac{4\BI}{1-\BI^2} kp\right)
 & \text{in Lor. case} \end{array}\right. \\
\label{pseudocaseq}
\epsilon_{IJKL} \left(\int_f \Sigma^{IJ}\right) \left( \int_f \Sigma^{KL}\right)
\hspace{-0.2cm}&=&\hspace{-0.2cm} \left\{ \begin{array}{ll} \tilde{C}_E \left(\frac{(j^+_f)^2}{(\BI + 1)^2}
- \frac{(j^-_f)^2}{(\BI - 1)^2}\right) &\hspace{-0.2cm}\text{in Eucl. case}\\
\tilde{C}_L
\left(k_f+\frac{1}{\BI}p_f\right)\left(k_f - \BI p_f\right)&\hspace{-0.2cm}\text{in Lor. case} \end{array}\right.
\hspace{1cm}\dummy
\end{eqnarray}
where $C_E, C_L, \tilde{C}_E, \tilde{C}_L$ are each a certain combination of the Planck length $\ell_\Pl$, $\BI$ and
numerical factors.
The integral $\int_{S} \Sigma^{IJ}$  of $\Sigma^{IJ}_{\mu\nu}(x)$ over a surface $S$ appearing in the above
equations is the standard
``differential form'' integral, defined by
\begin{equation}
\label{formint}
\int_S \Sigma^{IJ}
:= \int \Sigma_{\mu\nu}^{IJ}
\frac{\partial \tau^\mu}{\partial u} \frac{\partial \tau^\nu}{\partial v} du dv
\end{equation}
where $(u,v)$ are any choice of coordinates on $S$, and $x^\mu=\tau^\mu(u,v)$
is the four dimensional position of the point on $S$ with surface coordinates $(u,v)$.
(The result of the integral (\ref{formint}) is independent of the choice
 of $(u,v)$ and hence of $\tau^\mu(u,v)$).

\subsection{Simplicity and the LQG spin foam model}
\label{bfsimp_sect}

Recall the general strategy we are taking: One starts from the probability amplitude
$\hamp^{BF}(\mathcal{F}, \mathcal{L})$ for BF theory, and then restricts consideration to the case in which the BF
spin foam $(\mathcal{F}, \mathcal{L})$ satisfies some quantum version of the simplicity constraint
(\ref{simp}).
Just as the classical simplicity constraint is sufficient to recover classical gravity from BF theory, so too one expects
an appropriate quantum simplicity constraint to recover quantum gravity from quantum BF theory.

Different ways of imposing simplicity quantum mechanically then lead to different spin foam models of gravity.
We will present here only the most recent, commonly used way of imposing quantum simplicity, which
leads to the so-called `LQG spin foam model.'  This model is also variously referred to as the `EPRL', `EPRL-FK', or `EPRL-KKL' model,
after different authors who contributed to its development \cite{fk2007, elpr2007, kkl2009}.
At the center of this strategy of imposing simplicity is the so-called `linear simplicity constraint':
The condition that
\begin{equation}
\label{linsimp}
(G_{ve})^0{}_I \int_f \Sigma^{IJ} = 0
\end{equation}
for all $f$, $e$, and $v$ incident on one another.  It is called `linear' because it is linear
in the field $\Sigma_{\mu\nu}^{IJ}(x)$.
One can show that it implies $\Sigma_{\mu\nu}^{IJ}$ takes one of the following three forms:
\cite{engle2011a, engle2012}
\begin{eqnarray}
\label{pns_eq}
\text{\pns} && \Sigma_{\mu\nu}^{IJ}(x) =  \pm \left(e^I_\mu(x) e^J_\nu(x) - e^J_\mu(x) e^I_\nu(x) \right)
\text{ for some }e^I_\mu(x)\\
\label{degs_eq}
\text{\degs} && \Sigma_{\mu\nu}^{IJ}(x) \text{ is degenerate, that is, }
\epsilon_{IJKL} \epsilon^{\mu\nu\rho\sigma} \Sigma_{\mu\nu}^{IJ} \Sigma_{\rho\sigma}^{KL} = 0
\end{eqnarray}
Each of these constitutes a different sector of solutions to the equation (\ref{linsimp}); we have chosen the
symbols \pls, \negs, \degs {} to denote these sectors.
Notice that only sectors \pls, \negs yield a field $\Sigma$ of the form (\ref{simp}) required to obtain gravity.
In fact, as we will see later in the section on the semiclassical limit,
the existence of the last, degenerate, sector, will cause problems, and we will mention one way to
solve this problem.  (However, it should be noted that
the above three sectors of linear simplicity is already an improvement over the prior version of the simplicity
constraint used in the literature \cite{bc1997, bc1999, bhnr2004,engle2011}, which had five sectors.)

We just have discussed the \textit{classical} implications of the linear simplicity constraint (\ref{linsimp});
however, it is the \textit{quantum} implications for the BF spin foams that will yield us our quantum theory of
gravity.  From equations (\ref{leq}), (\ref{caseq}), and (\ref{pseudocaseq}) one can deduce
the consequences of linear simplicity (\ref{linsimp}) for the quantum numbers labeling the BF spin foams.
In the Euclidean case, these are precisely
\begin{displaymath}
j_f^\pm = \frac{1}{2}|1\pm \BI| j_{ef}
\end{displaymath}
and in the Lorentzian case,
\begin{displaymath}
\label{lorsimp}
p_f = \BI j_{ef} \text{ and } k_f = j_{ef},
\end{displaymath}
both remarkably simple forms.  These are the quantum simplicity constraints at the heart
of the LQG spin foam model of gravity.  After one imposes these constraints, one can ask:
What free spin foam labels are left?
In the Euclidean case, one starts out with the
BF spin foam labels $\{j_f^+,j_f^-, j_{ef}, n_{ef}\}$; the above quantum simplicity constraint
uniquely determines the labels $j_f^\pm$ in terms of $j_{ef}$, and furthermore forces that,
for each $f$, all the spins $j_{ef}$ are equal, whence we can write simply $j_f$.  Thus,
the remaining free labels are $\{j_f, n_{ef}\}$.  The same is true in the Lorentzian case: There one starts with
the labels $\{p_f, k_f, j_{ef}, n_{ef}\}$, simplicity determines $p_f$ and $k_f$ in terms of $j_{ef}$,
and all the spins $j_{ef}$ for a given face $f$ are equal, whence we may write $j_f$, and again the remaining free
labels are $\{j_f, n_{ef}\}$.  The key thing to note here is that \textit{in both cases, the remaining free labels
are exactly the same as the labels on the LQG spin foams introduced earlier}.
Thus, just as \textit{classically} the simplicity constraint reduces BF theory to gravity, so the quantum simplicity
constraint reduces BF spin foams to LQG spin foams.  That this key classical property is reproduced quantum mechanically
is one of the principal successes of the linear simplicity constraint as imposed in the LQG spin foam model,
and is what allows the LQG spin foam model to provide a dynamics for LQG, making it the first, and thus far only, spin foam model to
do so.  For other, more subtle, but no less interesting, arguments for this model,
we refer the reader to the original papers \cite{epr2007, fk2007, elpr2007, kkl2009}.

\subsection{Interpretation of LQG Spin foam quantum numbers: Quantum space-time geometry}
\label{qstgeom_sect}

LQG Spin foams describe the \textit{gravitational field}, and hence the \textit{geometry of space-time}.
We here take the time to explain how the labels of a LQG spin foam determine a \textit{discrete space-time geometry}.
This is important not only for understanding the meaning of the LQG spin foam labels, but will be central in
looking at the semiclassical limit of the resulting spin foam quantum theory --- that is, the limit in which
quantum mechanics is `turned off', and in which one should recover classical general relativity.

This section builds on sections \ref{sn_sect}-\ref{sndual_sect} on the discrete spatial geometry
determined by spin networks.  Just as in classical general relativity, where spatial geometry fits into the larger space-time
geometry, so too the quantum spatial geometry of spin networks fits consistently into the larger quantum space-time geometry of spin foams.

\subsubsection{Interpretation of quantum numbers and uniqueness of the space-time geometry}
\label{interpst_sect}

\begin{figure}
\begin{center}
\includegraphics[height=3in]{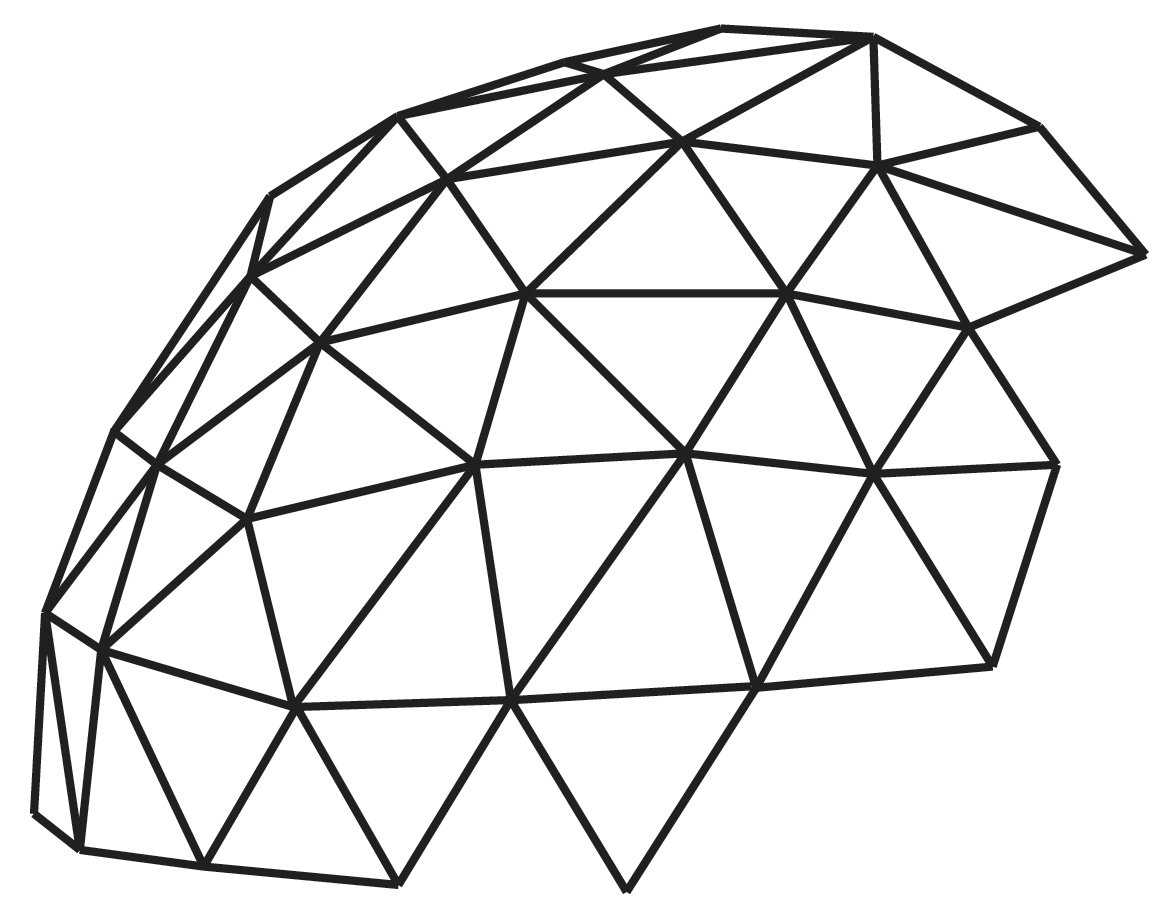}
\end{center}
\caption{\label{reggefig}
Illustration in two dimensions: Even though each 2-cell
is flat, when many are glued together, the resulting two dimensional cell
complex need not be flat, and in fact can approximate any curved geometry.
}
\end{figure}

The quantum numbers $\{j_f, n_{ef}\}$ determine a discrete space-time geometry by determining uniquely
the geometry of \textit{each 3-cell} $e\dual$ in the dual complex $\mathcal{F}\dual$.  Specifically,
for each 3-cell $e\dual$, the area of each face $f\dual$ of $e\dual$ is equal to
$8\pi \ell_\Pl^2 \BI \sqrt{j_f(j_f+1)}$, and the interior angle
$\theta[e\dual, f\dual, \tilde{f}\dual]$ between each pair of faces $f\dual, \tilde{f}\dual$
within $e\dual$ is given by the equation
\begin{displaymath}
\cos \left(\theta[e\dual, f\dual, \tilde{f}\dual]\right) = - n_{ef} \cdot n_{e\tilde{f}} .
\end{displaymath}
These areas and angles are precisely the same areas and angles used to interpret the LQG spin-network
labels in section \ref{sndual_sect}, except now in a four dimensional context.
Using a theorem by Minkowski \cite{minkowski1897}, one can show \cite{bds2010}
that these areas and angles are sufficient to uniquely determine a flat geometry within each three-cell $e\dual$.
Because one is now working in four dimensions, each 3-cell is now part of the boundary of two \textit{4-cells}.
The geometry of all the 3-cells $e\dual$ bounding each 4-cell $v\dual$
is sufficient to determine a flat geometry within each 4-cell \cite{bdfgh2009, connelly1993}.
By determining the geometry within each 4-cell,
one determines a geometry of the entire space-time.
This geometry is \textit{piece-wise flat}: Within each 4-cell it is flat, but the resulting overall geometry of the larger space-time
certainly need not be flat, and indeed can approximate any desired space-time geometry arbitrarily well.
(See figure \ref{reggefig} for a depiction of this phenomenon in two dimensions.)
This is the discrete, quantum space-time
geometry determined by a loop quantum gravity spin foam $(\mathcal{F}, \{j_f, n_{ef}\})$.

\subsubsection{Existence}
\label{exist_sect}

The above discussion explains how the quantum numbers $\{j_f, n_{ef}\} $ of a LQG spin foam are sufficient to
uniquely determine a piecewise flat space-time geometry, \textit{assuming} there \textit{exists} a space-time geometry
compatible with the given spin foam data.  This will not always be the case:  there are constraints on the spin-foam data.
Specifically, in order for a compatible space-time
geometry to exist, two constraints must be satisfied \cite{bdfgh2009, bdfhp2009}: the \textit{closure} and
\textit{gluing} constraints \cite{ds2008, dr2008}.
The closure constraint requires that, in each dual 3-cell $e\dual$, one has
\begin{displaymath}
\sum_{f\text{ incident at }e} j_f n_{ef}^i = 0 .
\end{displaymath}
By the same theorem of Minkowski cited earlier
%
%
\cite{minkowski1897, bds2010}, this constraint is sufficient
to ensure that a consistent geometry for the \textit{3-cell} $e\dual$ can be reconstructed.
The second constraint, the \textit{gluing} constriant, is the requirement that all of these 3-cell geometries
\textit{fit} together consistently --- that is, when two 3-cells share a face, they should have the same area and shape.
When this is true, then the \textit{4-cell} geometries will also exist and fit together, yielding a full piecewise flat
space-time geometry consistent with the given spin foam data.

\subsection{The loop-quantum-gravity spin foam amplitude}

We are now ready to implement the last step of the strategy to define a spin foam model of quantum gravity.
We have already described BF theory and its quantum histories in subsections \ref{BFgravity_sect}-\ref{sfbf_sect},
and have introduced a way of imposing the simplicity constraint quantum mechanically in subsection \ref{bfsimp_sect}.
Just as the set of classical BF fields satisfying classical simplicity coincides with the fields describing
classical gravity, so too, we have seen that the set of BF spin foams satisfying quantum
simplicity as presented above is in 1-1 correspondence with loop quantum gravity spin foams.
Let $\mathcal{I}$ denote the 1-1 map from LQG spin foam labels on a given
$\mathcal{F}$ to BF spin foam labels satisfying simplicity on the same
$\mathcal{F}$.
The last step of the strategy is to restrict the spin foam amplitude  $\hamp^{BF}(\mathcal{F},\mathcal{L})$
of BF theory to spin-foams satisfying simplicity.
This leads one to assign the following probability amplitude to each LQG spin foam
$(\mathcal{F}, \{j_f, n_{ef}\})$:
\begin{equation}
\label{lqgamp}
\hamplqg(\mathcal{F}, \{j_f, n_{ef}\}) := \hamp^{BF}(\mathcal{F}, \mathcal{I}(\{j_f, n_{ef}\}))
\end{equation}
This is the LQG spin foam amplitude, defining the LQG spin foam model of quantum gravity.
It exists in both Euclidean and Lorentzian versions, depending on which BF theory one starts with.
$\hamplqg(\mathcal{F}, \{j_f, n_{ef}\})$ gives the probability amplitude for the single quantum space-time history
$(\mathcal{F}, \{j_f, n_{ef}\})$, the geometrical meaning of which has been explained in the previous section.
This is the principal spin foam amplitude that will be used in the rest of this chapter.

Note that while the BF spin foam amplitude is a well-established result of an exactly soluble theory,
the LQG spin foam amplitude (\ref{lqgamp}) must be considered a proposal due to the non-trivial
decision involved in the way the simplicity constraint is imposed.  Nevertheless, the particular way of
imposing the simplicity constraint presented above has compelling properties \cite{elpr2007, dr2009},
especially the exact reduction of BF spin foam labels to those of LQG,
which no other strategy thus far has.

\section{Regge action and the semiclassical limit}

We turn attention now to the classical limit of the LQG spin foam model.
Recall from section \ref{qhist_sect} that the classical limit is defined as the limit in which appropriate combinations of
physical quantities become large compared with Planck's constant. In the case of gravity, the relevant physical quantities
are \textit{geometrical} and have dimensions of some power of length.
As mentioned earlier in this chapter,  there is a unique combination of Newton's gravitational constant, Planck's constant
divided by $2\pi$, and the speed of light, with dimensions of length: the Planck length, $\ell_\Pl = \sqrt{\frac{G \hbar}{c^3}}$.
The classical limit of a quantum theory of gravity arises
when geometrical quantities become large compared to the corresponding power of the Planck length. In this limit, quantum theory can be neglected,
and, in order for the theory to remain compatible with the many successful experimental and observational tests of general relativity, it is necessary for the theory, in this limit, to become general relativity.

Recall from section \ref{qhist_sect} that, when a quantum theory is formulated in terms of a path integral,
the form $e^{iS}$ of the amplitude for individual histories is important not only to have equivalence with
the canonical quantum dynamics, but also important for ensuring the correct classical limit of the theory.

This leads us to ask: Is the LQG spin foam amplitude, derived above, equal to $e^{iS}$, with $S$ an appropriate action
for gravity?
Except for one subtle point to be discussed further at the end of this subsection, this question has been answered
in the affirmative in the limit in which geometrical quantities are large compared to the Planck scale,
and in the special case in which the dual cell complex $\mathcal{F}\dual$ consists in cells of the simplest type,
called \textit{simplices}.
The limit in which geometrical quantities are large compared to the Planck scale is of course just the classical limit.
However, because the probability amplitude for individual histories is a fundamentally quantum mechanical
object with no classical analogue, one usually instead refers to this as the \textit{semiclassical limit} of the amplitude.
When all cells of $\mathcal{F}\dual$ are simplices, $\mathcal{F}\dual$ is called a
\textit{simplicial complex}. The possible piece-wise flat geometries on such a complex are called
\textit{Regge geometries}.  Before stating the semiclassical limit of the LQG spin foam amplitude,
we explain in more detail simplicial complexes and Regge geometries.

\subsection{Regge geometries}
\label{regge_sect}

\begin{figure}
\begin{center}
\begin{minipage}[b]{0.7in}
\includegraphics[height=0.5in]{dual13.jpg}
\\
\vspace{0cm}\dummy
\end{minipage}
~\includegraphics[height=0.9in]{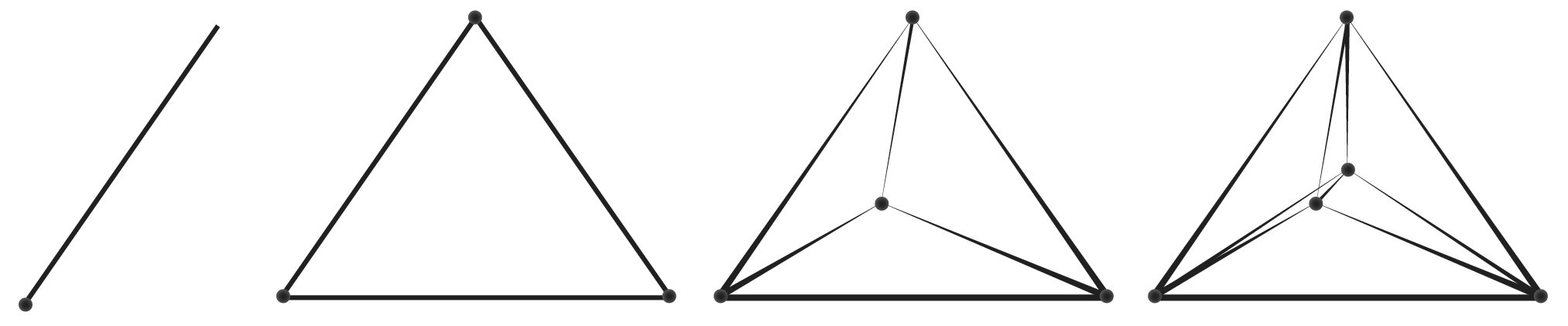}
\end{center}
\caption{\label{simplicesfig}
0-simplex, 1-simplex, 2-simplex, 3-simplex, and a 4-simplex, projected into
a three-dimensional plane for visualization.
}
\end{figure}

Until this point we have spoken of general \textit{cells} in the dual cell complex $\mathcal{F}\dual$.
In each dimension $n$, there is a certain type of simplest possible cell called a \textit{simplex}, plural \textit{simplices}.
When one wishes to specify the dimension $n$ of a simplex, one uses the term \textit{$n$-simplex}.
0-simplices are \textit{points}, 1-simplices are \textit{line segments}, 2-simplices are \textit{triangles},
and 3-simplices are \textit{tetrahedra}. In four dimensions, there is no common term for the simplest possible cell; it is
therefore simply called a 4-simplex.  (See figure \ref{simplicesfig}.)

Recall from section \ref{qstgeom_sect}
how each LQG spin foam $(\mathcal{F}, \{j_f, n_{ef}\})$ determines a space-time geometry which is
\textit{flat} in each dual 4-cell $v\dual$ --- that is, a \textit{piece-wise flat space-time geometry}.
When we furthermore require that all of the dual 4-cells be \textit{simplicial}, the resulting
geometry is called a \textit{Regge geometry}.
Regge geometries were first introduced by Tullio Regge \cite{regge1961}
and are well studied in the literature.
Usually Regge geometries are specified by giving the lengths of all 1-simplices, as this information is equivalent to
specifying a piecewise flat geometry, as noted in Regge's original paper.
%
%
We furthermore note that the Regge geometries determined by spin-foams are slightly more restricted than what is usual for Regge calculus, in that the areas of triangles are restricted to belong to the canonical area spectrum given in equation (\ref{linkarea}).

When one takes the standard action for general relativity, the Einstein-Hilbert action,
and evaluates it on Regge geometries, one obtains the \textit{Regge action} \cite{fl1984}.  Thus, in order to ensure that our spin foam model of
gravity has the correct classical limit (namely, general relativity), one would like the amplitude for such a spin foam to be
\begin{displaymath}
\hamp(\mathcal{F}, \{j_f, n_{ef}\}) = (\text{positive real number}) \exp\left(i S_{\Regge} \right)
\end{displaymath}
where $S_\Regge$ is the Regge action.

\subsection{Semiclassical limit}
\label{semicl_sect}

With the above background, we are ready to state the result on the semiclassical limit
of the LQG spin foam amplitude.  The LQG spin foam amplitude follows the
spin foam Ansatz (\ref{sfamp}), so that it decomposes into face, edge, and vertex amplitudes:
\begin{equation}
\label{lqgsfamp}
\hamplqg(\mathcal{F}, \{j_f, n_{ef}\})
= \left(\prod_{f \in \mathcal{F}} \hamplqg_f\right)
\left(\prod_{e \in \mathcal{F}} \hamplqg_e\right)
\left(\prod_{v \in \mathcal{F}} \hamplqg_v\right).
\end{equation}
Recall every complex number $A$ can be decomposed as $A = |A|e^{i\theta}$ where $\theta$ is the \textit{phase}.
Following the argument of section \ref{qhist_sect}, in order to obtain the correct
classical limit of the quantum theory, it is sufficient for the \textit{phase} of the amplitude to become the classical action.
The face and edge amplitudes are real. Hence they can contribute at most an integer multiple of $\pi$ to the phase of the full amplitude.
The interesting contribution to the phase of the amplitude will thus be from the vertex amplitudes, and so
one focuses on the semiclassical limit of primarily the vertex amplitudes.

Each vertex amplitude $\hamplqg_v$ can literally be understood
as the spin foam amplitude for a single 4-cell, $v\dual$.
As mentioned above, the semiclassical limit has thus far only
been carried out for the case in which each 4-cell is a \textit{4-simplex}, and thus we
restrict consideration to the case in which $v\dual$ is a 4-simplex.
In this case, the vertex amplitude $\hamplqg_v$ depends on the 10 spins $j_f$ on the 10 faces incident at $v$, and on the 20 vectors $n_{ef}$ labeling the 5 edges $e$ incident at $v$ and the 4 faces $f$ incident at each of these 5 edges. To emphasize this
dependence, we write $\hamplqg_v(\{j_f, n_{ef}\})$, where it is understood that there is only dependence
on these spins and vectors.

Recall that the semiclassical is the limit in which geometric quantities become large compared to the Planck scale.
All geometric quantities determined by the labels $\{j_f, n_{ef}\}$ scale directly with the spins $j_f$, so that an easy way to take the
semiclassical limit  is to rescale all of these spins by some common parameter $\lambda$, and then
take the limit in which $\lambda$ becomes large.  Thus, concretely, to look at the semiclassical limit of the vertex amplitude,
one looks at the limit of $\hamplqg_v(\{\lambda j_f, n_{ef}\})$ as $\lambda$ becomes large.

The form of the semiclassical limit of $\hamplqg_v(\{\lambda j_f, n_{ef}\})$ is different depending on
the type of geometry, or lack thereof, determined by the labels $\{j_f, n_{ef}\}$.
The form of the semiclassical limit of the Euclidean version of the vertex amplitude falls into 3 different cases, whereas
that of the Lorentzian version is more subtle and falls into 4 different cases.  Because the Euclidean case is sufficient to demonstrate
the important issues, and is simpler, we restrict the following presentation to the Euclidean case.
In section \ref{qstgeom_sect} we reviewed how, if the spin foam labels $\{j_f, n_{ef}\}$ satisfy what are called the
closure and the gluing constraints, then they uniquely determine a (possibly degenerate) flat geometry for
the 4-cell $v\dual$ (which in this case is a 4-simplex).   If the tetrahedra on the boundary of $v\dual$,
as determined by this geometry, all have non-zero volume, we say that the labels determine a
\textit{non-degenerate boundary geometry} of $v\dual$.  If any of the tetrahedra have zero volume, we
say that the labels $\{j_f, n_{ef}\}$ determine a \textit{degenerate boundary geometry}.
If the the labels do not satisfy the closure and gluing constraints, we say that they are \textit{non-geometric}.
We give below the semiclassical
limit, which we denote by the symbol `$\sim$', of the Euclidean version of $\hamplqg_v$ in each these three different cases.
\begin{enumerate}
\item
For labels determining a non-degenerate boundary geometry,
\begin{equation}
\label{nondeglim}
\hamplqg_v(\{\lambda j_f, n_{ef}\}) \sim
\lambda^{-12}\left(C_1 e^{iS_\Regge} + C_2 e^{-iS_\Regge} + C_3 e^{\frac{i}{\BI}S_\Regge} + C_4 e^{-\frac{i}{\BI}S_\Regge}\right).
\end{equation}
\item
For labels determining a degenerate boundary geometry,
\begin{equation}
\label{deglim}
\hamplqg_v(\{\lambda j_f, n_{ef}\}) \sim
 \lambda^{-12} C.
\end{equation}
\item
For non-geometric labels, the probability amplitude
$\hamplqg_v(\{\lambda j_f, n_{ef}\})$ decays exponentially with $\lambda$, that is, as fast as $e^{-\lambda}$,
so that such labels are `suppressed' by the vertex amplitude.
\end{enumerate}
In the above formulae, $C_1,C_2,C_3,C_4$, and $C$ are independent of $\lambda$.
Note that the only labels not suppressed are the ones that actually correspond to piece-wise flat (possibly degenerate) space-time
geometries.
In addition to this, the fact that the exponential of $i$ times various multiples of the action appears in the semiclassical limit of the vertex amplitude is encouraging.  However, this is not yet sufficient to ensure the correct classical limit: Not only are
some unphysical, degenerate geometries not suppressed, but even for the non-degenerate geometries, the asymptotic amplitude
(\ref{nondeglim}) is not yet the Feynman amplitude.  The Feynman amplitude would consist in only the first term in
(\ref{nondeglim}). There is a reason for the non-suppression of the degenerate configurations in (\ref{deglim}) 
as well as the extra terms in (\ref{nondeglim}); in a moment, we will remark on this reason,
as well as mention a solution to the problem.
That these extra terms spoil the classical limit of the theory can be seen by looking at spin foams on triangulations with more
than one 4-simplex.
In this case, even if we assume that the geometries of all 4-simplices are non-degenerate, 
one still has the four terms in (\ref{nondeglim}) \textit{for each 4-simplex}. 
When these four terms are substituted into the expression (\ref{lqgsfamp}) for the full amplitude, 
one obtains cross-terms.
Each of these cross-terms is equal to the exponential of a sum of terms, one for each 4-simplex, equal to the Regge
action for that 4-simplex times differing coefficients, yielding what can be called a `generalized Regge action' \cite{engle2012,
mp2011, hz2011}.  The extrema of this `generalized Regge action' are \textit{not} the Regge equations
of motion and hence \textit{not} those of general relativity, so that general relativity fails to be recovered in
the classical limit.

As shown in the recent work \cite{engle2011}, the extra terms causing this problem are due precisely
to the presence of the multiple sectors of solutions to the simplicity constraint presented in section \ref{BFgravity_sect}, as well as
the presence of different ``orientations'' as dynamically determined by the co-tetrad field $e_\mu^I$.
Once these sectors and orientations are properly handled \cite{engle2011a, engle2012}, one arrives at what is called the \textit{proper loop quantum gravity vertex amplitude}.  Its semiclassical limit includes only the single term consisting in the exponential of $i$ times the classical action, 
\begin{displaymath}
\hamp_v^{\text{(+)}}(\{\lambda j_f, n_{ef}\}) \sim
\lambda^{-12}C_1 e^{iS_\Regge},
\end{displaymath}
thereby solving the above problem and giving reason to believe
that the resulting spin-foam model will yield a correct classical limit.

\section{Two-point correlation function from spin foams}

In this section we review a calculation in spin foams which has played an important role in the development of the field:
The calculation of the two point correlation function of quantum gravity.
The two point correlation function of a quantum field theory
is the simplest quantity one can calculate which directly probes the ``non-classical-ness'' of the theory and thus provides one with
a genuinely quantum mechanical prediction.
In order to set the stage for this calculation, we begin with a technical discussion of how one sums over spin foams.

\subsection{The complete sum over spin foams}
\label{sfsum_sect}

Let us first recall how the spin foam amplitude
discussed in the last few sections fits into the overall calculation scheme.  As discussed in section \ref{bkgr_sect},
the amplitude for a given gravitational history
is used to calculate the \textit{probability amplitude} for a \textit{canonical quantum state} on the boundary of a given space-time region.
In the case where this canonical quantum state is an eigenstate $|h \rangle$ of spatial geometry on the boundary
of some space-time region $R$, the probability amplitude takes the form (\ref{gampeq})
\begin{equation}
\label{formalamp}
\gravamp(|h\rangle, R)
= \int_{g|_{\partial R} = h}  e^{iS[g]} \mathcal{D} g .
\end{equation}
Spin foams provides a way to make the above formal prescription concrete, by using the lessons of loop quantum gravity.
Loop quantum gravity tells us that the correct eigenstates of spatial geometry on $\partial R$ are the
\textit{spin network states} $|\gamma, \{j_\link, n_{\nu\link}\}\rangle$,
labeled by a graph $\gamma$ on the boundary of $R$, spins $j_\link$ and unit 3-vectors $n_{\nu\link}$ as in section
\ref{sn_sect}. The integral over continuum geometries is then replaces by the sum over discrete space-time geometries represented
by spin foams.  The formal expression (\ref{formalamp}) is then replaced by the concrete spin foam expression
\begin{displaymath}
\hamplqg(|\gamma, \{j_\link, n_{\nu\link}\}\rangle, R)
:= \sum_{\substack{\mathcal{F}\text{ such that }\\ \mathcal{F}\cap\partial R = \gamma}}
\sum_{\substack{\{j_f, n_{ef}\}\text{ such that }\\ \{j_f, n_{ef}\}|_{\partial R} = \{j_\link, n_{\nu\link}\}}}
\hamplqg(\mathcal{F}, \{j_f, n_{ef}\})
\end{displaymath}
Before using this expression, there is still one more issue that must be addressed.
Two different types of apparently infinite sums appear in the above expression:
(1.) the sum over possible \textit{two-complexes} $\mathcal{F}$ and (2.) the sum over possible \textit{labels} $\{j_f, n_{ef}\}$
on the two complex.
There is an infinite number of two-complexes, giving rise to the first potential source of infinity, and, for each two-complex,
there are an infinite number of possible ways to label it, giving rise to the second potential source of infinity.

To aide in addressing the first of these potential infinities, it is useful to separate the sum over two-complexes
into first a sum over \textit{numbers of vertices} $N$, and then a sum over two-complexes with $N$ vertices:
\begin{displaymath}
\hamplqg(|\gamma, \{j_\link, n_{\nu\link}\}\rangle, R)
:= \sum_{N=0}^\infty \quad \sum_{\substack{\mathcal{F}\text{ with }N\text{ vertices,}\\ \text{such that}\\ \mathcal{F}\cap\partial R = \gamma}}
\sum_{\substack{\{j_f, n_{ef}\}\text{ such that }\\ \{j_f, n_{ef}\}|_{\partial R} = \{j_\link, n_{\nu\link}\}}}
\hamplqg(\mathcal{F}, \{j_f, n_{ef}\}).
\end{displaymath}
For each $N$ there are only a finite number of two-complexes, so the the potential infinity resides only in the sum over $N$.
To handle this, one usually introduces a small, positive real number $\lambda$, raised to the power $N$:
\begin{equation}
\label{lambdaamp}
\hamplqg(|\gamma, \{j_\link, n_{\nu\link}\}\rangle, R)
:= \sum_{N=0}^\infty \lambda^{N} \sum_{\substack{\mathcal{F}\text{ with }N\text{ vertices,}\\ \text{such that}\\ \mathcal{F}\cap\partial R = \gamma}}
\sum_{\substack{\{j_f, n_{ef}\}\text{ such that }\\ \{j_f, n_{ef}\}|_{\partial R} = \{j_\link, n_{\nu\link}\}}}
\hamplqg(\mathcal{F}, \{j_f, n_{ef}\}).
\end{equation}
This has the effect of making each consecutive term in the sum over $N$ smaller, and so ensuring convergence.
The insertion of the power of $\lambda$ not only brings this potential infinity under control, it also allows the resulting spin foam theory
to be recast in terms of something called a \textit{group field theory} \cite{dfkr1999, oriti2009, ggr2010},
thereby enabling a wide array of developed tools to be used in the study of the theory.

The second potential infinity comes from the sum over \textit{labels} on each two-complex.
There are indications \cite{clrrr2012} that the proper vertex \cite{engle2011a, engle2012} introduced in section
\ref{semicl_sect} may solve this second problem, though, at the moment, these are only indications.
Other promising research directions related to this question include \cite{prs2008, fl2002, fl2004}.
However, in the following application, we look only at the terms in the sum (\ref{lambdaamp})
with the lowest power of $\lambda$.  One can show that the sum over labels for the lowest power of $\lambda$ is finite,
so that, at least for the calculations considered below, this second infinity is not an issue.

\subsection{The calculation}

To define the two-point correlation function, we first
introduce the idea of the \textit{expectation value} of an operator $\hat{O}$ in a given boundary state $\Psi$,
as computed using the path integral formalism for some region $R$ of space-time.
The expectation value is the \textit{average result} one would obtain by measuring the quantity $\hat{O}$ when the
system is in the state $\Psi$.  It is denoted $\langle \hat{O} \rangle_\Psi$ and is given by the expression
\begin{displaymath}
\langle \hat{O} \rangle_\Psi := \frac{\tamp(\hat{O}\Psi,R)}{\tamp(\Psi,R)}.
\end{displaymath}

For illustrative purposes, let us first consider the case of a scalar field theory.
In this case, as an aside for those more familiar with standard quantum field theory, when $\Psi$ is an eigenstate $|\varphi(\spp{x})\rangle$ of the field
operator $\hat{\varphi}(\spp{x})$ and $\hat{O}$ is a function $O(\hat{\varphi}(\spp{x}))$ of the field operator,
the above expression for the expectation value takes the more familiar form
\begin{displaymath}
\langle \hat{O} \rangle_{\Psi} :=  \frac{\int_{\phi|_{\partial R} = \varphi} O(\phi) e^{iS[\phi]} \mathcal{D} \phi }
{\int_{\phi|_{\partial R} = \varphi} e^{iS[\phi]} \mathcal{D} \phi } .
\end{displaymath}
Given a canonical quantum state $\Psi$,
and any two points $\spp{x}$ and $\spp{y}$ on $\partial R$,
the two-point correlation function is defined as
\begin{displaymath}
D_\Psi (\spp{x},\spp{y}) :=
\langle \hat{\varphi}(\spp{x}) \hat{\varphi}(\spp{y}) \rangle_\Psi
- \langle \hat{\varphi}(\spp{x}) \rangle_\Psi
\langle \hat{\varphi}(\spp{y}) \rangle_\Psi .
\end{displaymath}
In the classical theory, the state of the system is, and therefore uniquely determines, the value of the field $\varphi(\spp{x})$
and its conjugate momentum.  Thus, classically, given the state of the system,
the outcome of a measurement of $\varphi(\spp{x})$ is certain, so that the expectation value of
$\varphi(\spp{x})$ is just $\varphi(\spp{x})$, and the expectation value of $\varphi(\spp{x})\varphi(\spp{y})$
is just $\varphi(\spp{x})\varphi(\spp{y})$, so that  the classical two-point correlation function is just zero.
Its deviation from zero can therefore be thought of as a measure of the `non-classical-ness' of the theory, providing us with an
essentially quantum mechanical prediction of the theory.

\begin{figure}
\begin{center}
\includegraphics[height=2in]{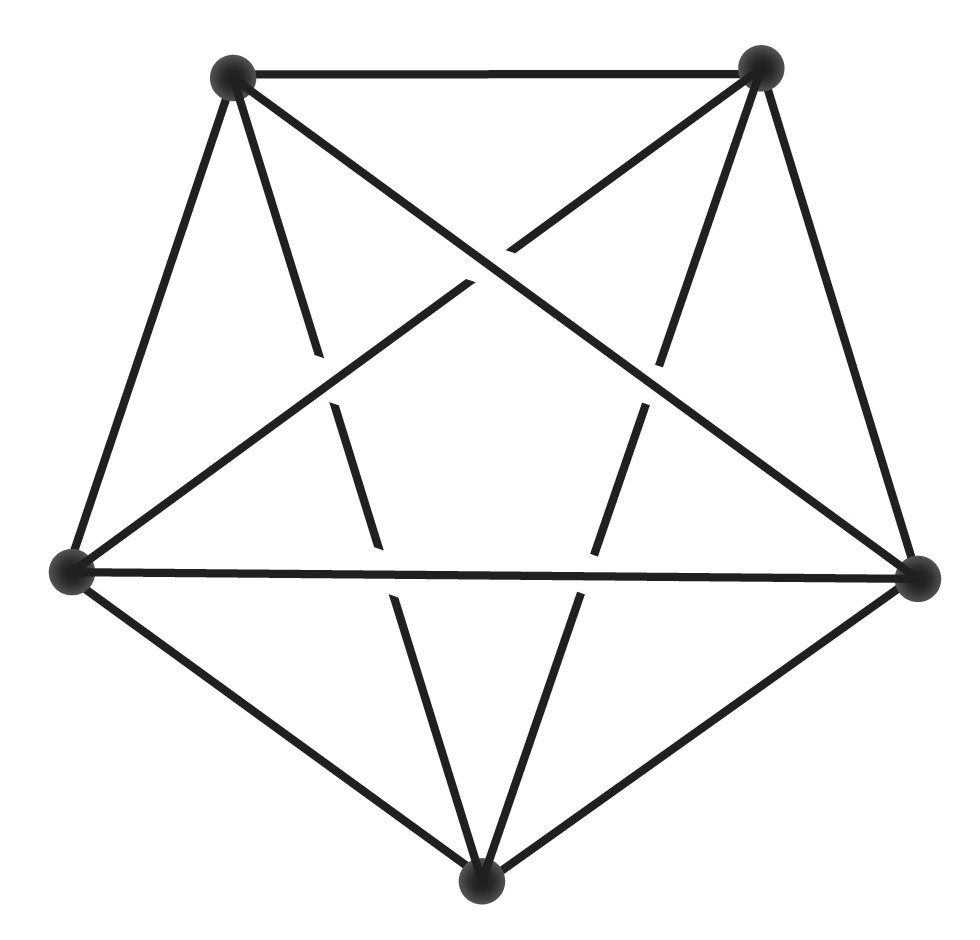}
\end{center}
\caption{\label{simplexgraphfig}
The connectivity of the graph $\gamma$ used for the boundary state.
}
\end{figure}

The two-point correlation function for loop quantum gravity, as determined by the loop quantum gravity spin foam model,
has been calculated in the works  \cite{rovelli2005, ar2007, ar2007a, bmp2009b, bd2011},
for both the Euclidean and the Lorentzian versions of the model.
The \textit{field operator} in this case (that is, the operator playing the role of $\hat{\varphi}(\spp{x})$ above), is the
\textit{metric tensor} field operator, which takes a discrete form in the case of loop quantum gravity.
The boundary state $\Psi$ that is considered is a linear combination of spin networks
based on a fixed graph $\gamma$ on the boundary of $R$ having the structure indicated in
figure \ref{simplexgraphfig}.

In order to describe precisely the metric tensor operator and the way it is discretized when
acting on the state $\Psi$,
we use again the notion of the complex $\gamma\dual$ dual to $\gamma$ within the three dimensional
boundary of $R$.
The graph $\gamma$ within $\partial R$ consists in nodes and links.
The dual to each node $\node$ in $\gamma$ is a three-dimensional region (a 3-cell) $\node\dual$,
and the dual to each link $\link$  incident at $\node$
is a two-dimensional surface (a 2-cell) $\link\dual$ in the boundary of $\node\dual$.
For the specific case of the graph
$\gamma$ given in figure \ref{simplexgraphfig}, each 3-cell $\node\dual$  is a \textit{tetrahedron} and each 2-cell $\link\dual$ is a \textit{triangle},
so that the dual cell  complex $\gamma\dual$ is a \textit{simplicial} complex in the same sense as section \ref{regge_sect}, but one
dimension lower, and so provides a \textit{triangulation} of the boundary of $R$.  In fact, $\gamma\dual$ is the boundary of a single four-simplex (see section \ref{regge_sect}); this will be important in a moment.

In terms of the dual cell complex $\gamma\dual$, the metric tensor operator acting on $\Psi$
is defined as follows.
Classically, the metric tensor determines areas and angles.
In LQG one has operators corresponding to the areas of the triangles $\link\dual$
and the interior angles of the tetrahedra $\node\dual$ in $\gamma\dual$.
One can assemble these operators into a single `metric tensor-like matrix' as
\begin{equation}
\label{metricop}
\hat{h}^{\link \link'}(\node) := \hat{A}_{\link\dual} \hat{A}_{{\link'}\dual} \cos(\hat{\theta}[\node\dual; \link\dual, {\link'}\dual])
\end{equation}
Where $\hat{A}_{\link\dual}$ is the area operator for the triangle $\link\dual$, and
$\hat{\theta}[\node\dual; \link\dual, {\link'}\dual]$ is the interior angle between the triangles $\link\dual$, ${\link'}\dual$
within the tetrahedron $\node\dual$.
These area and angles are precisely the same areas and angles used to interpret
LQG spin networks in section \ref{sn_sect}-\ref{sndual_sect} and LQG spin foam labels in section \ref{interpst_sect}, 
except now cast as operators.

We have mentioned that we chose the boundary state $\Psi$ to be based on the graph $\gamma$ in $\partial R$.
We furthermore choose it to be a
\textit{coherent state}, that is, a quantum state which approximates as well as possible a particular \textit{classical state}
--- one says it is `peaked' on a particular classical state.
A classical state in this case consists in an intrinsic geometry of the boundary, described by a matrix $\mathring{h}$ of areas and angles
as in (\ref{metricop}), \textit{together with} a specification $\mathring{\Pi}$ of its conjugate momentum, which,
as mentioned in section \ref{gengrav_sect},
describes how $\partial R$ bends in the larger, four-dimensional space $R$. That is, $\mathring{\Pi}$ describes the
\textit{extrinsic} geometry of $\partial R$.
The state $\Psi$ which is used in the calculation is peaked on a particular $\mathring{h}$ and $\mathring{\Pi}$
which are chosen to be as simple as possible, namely they are chosen to describe the intrinsic and extrinsic geometry of the boundary of a
regular 4-simplex --- that is, a 4-simplex where all of the edges are of equal length.

Now that the nature of the metric tensor field operator and the boundary state have been clarified, we return to the expression for
the two-point correlation function.  Again, one first defines the notion of the expectation value of a given operator $\hat{O}$
in the boundary state $\Psi$:
\begin{displaymath}
\langle \hat{O} \rangle_{\Psi} := \frac{\tamplqg(\hat{O}\Psi, R)}{\tamplqg(\Psi, R)}
\end{displaymath}
and the two-point correlation function of the metric tensor operator (\ref{metricop}) is then
\begin{equation}
\label{lqgcorr}
G^{\link_1 \link_1' \link_2 \link_2'}(\node_1, \node_2)
:= \langle \hat{h}^{\link_1 \link_1'}(\node_1)  \hat{h}^{\link_2 \link_2'}(\node_2)  \rangle_\Psi
- \langle \hat{h}^{\link_1 \link_1'}(\node_1) \rangle_\Psi \langle \hat{h}^{\link_2 \link_2'}(\node_2)  \rangle_\Psi .
\end{equation}
One can expand this quantity in a power series in the coupling constant $\lambda$ introduced in section \ref{sfsum_sect}, and
what has been so far calculated is the lowest order term in this series, which corresponds to summing over spin foams
which include only a single vertex, and
which, on the boundary of $R$, coincide with the graph $\gamma$.
For the given graph $\gamma$ in figure \ref{simplexgraphfig}, there is only one such spin foam, and its one vertex is dual (in four dimensions)
to a single 4-simplex, of which $\gamma\dual$ forms the boundary.

The quantity (\ref{lqgcorr}) has been calculated in \cite{bmp2009b, bd2011} (to leading order in $\lambda$).
It has been found to match, at least in part, the same result one would calculate
in a more classic, but incomplete quantum gravity framework --- linearized quantum gravity
\cite{veltman1975, burgess2003, ar2007a}
--- the beginnings of which date back to the work of Rosenfeld, Fierz and Pauli in the 1930's\cite{rovelli2004}.
Linearized gravity is a simplified version of gravity obtained by assuming that space-time geometry is close to flat, so
that the metric tensor $g_{\mu\nu}$ is equal to a flat background metric $\eta_{\mu\nu}$ plus some small change
$\varepsilon h_{\mu\nu}$ where the components of $h_{\mu\nu}$ are of order one, while $\varepsilon$ is much less than one.
If one substitutes $g = \eta + \varepsilon h$ in to the standard action of gravity $S^\text{grav}[\eta + \varepsilon h]$, one can then expand the
action in powers of $\varepsilon$.  The term with the lowest power of $\varepsilon$, in this case 2, is then the action for linearized gravity.
%
%
Because the linearized action involves only first and second powers of the basic variable of the theory (usually taken to be $h$),
the theory can be exactly quantized.
 The two-point correlation function (\ref{lqgcorr}) calculated using the LQG spin foam model differs
from the two-point correlation function of linearized quantum gravity by addition of a term
which goes to zero as the Barbero-Immirzi parmeter $\BI$ goes to zero.  This extra term thus yields a new signature of
the loop quantum gravity spin foam dynamics; its significance has yet to be fully understood.
%
%

\section{Discussion}

In the spin foam approach to quantum gravity, one uses what has been learned from canonical loop quantum gravity about quantum \textit{space}
to construct a path integral approach to quantum gravity in which one sums over \textit{quantum space-times}.
The resulting framework  allows for simpler concrete calculations of the consequences
of dynamics than was possible using the canonical methods of loop quantum gravity alone --- we have seen this already above
in the calculation of the two point correlation function, and one can also see it in the first steps of the application of the full
spin foam theory to cosmology \cite{rv2009, rv2008, brv2010, vidotto2011},
a topic which we have not been able to discuss in this chapter.

Beyond these basic developments, the spin foam approach to quantum gravity has raised
other interesting questions and led to further lines of research which are ongoing.
These include among others work on how the theory appears on different length scales
(so-called \textit{renormalization} of spin foams)
\cite{rs2010, dem2011, bds2011, rivasseau2011, co2012}, systematic issues in the derivation of spin foams
\cite{dr2009, eht2009, han2009, brr2010, dr2010}, mathematical tools and equivalent reformulations
of spin foam theory \cite{bhkkl2010, dfl2011, dl2011},
inclusion of matter \cite{bhrwm2010, hr2011, bkrv2011},
the relation between the dynamics defined by spin foam sums and the dynamics defined by the
Hamiltonian constraint in loop quantum gravity \cite{ar2010, atz2011},
and the surprising relation of spin foams to other approaches to quantum gravity, specifically noncommutative geometry
\cite{fl2005, fl2005a} and group field theory \cite{dfkr1999, oriti2009, ggr2010}.
These are only a few representative works of the various research directions inspired by spin foams.

If one is to distill a single lesson from the spin foam program, it is perhaps this: In constructing a path integral formulation of a quantum theory, it is important to remember the role played by canonical quantization in determining the \textit{potentially
discrete nature of the histories one sums over}.
A proper path integral approach to quantum gravity, strictly speaking,
should not define transition amplitudes between classical geometries, but rather between
\textit{canonical quantum states of quantum gravity}, and one should not sum over classical space-time geometries, but rather
\textit{histories of quantum states}.  This is what leads directly to the spin foam program.
In addition to having this firm theoretical basis, the final framework provides a way to combine the advantages of canonical quantum gravity, with its predictions of
discrete geometry, black hole entropy, and quantum cosmology, with the
manifest unity of space and time made possible by the
path integral approach, a unity of space and time at the heart of both special and general relativity.
Lastly, in addition to these theoretical and aesthetic advantages, as we have already touched upon above, the resulting framework allows  for simple, concrete calculations involving dynamics by offering an alternative to the
task of finding general solutions to the quantum Hamiltonian constraint.

\section*{Acknowledgements}
The author thanks his wife, Sabine Engle, for careful assistance with the figures in this chapter, 
the editors and an anonymous referee for assistance in improving the chapter,
and Christopher Beetle for pointing out reference \cite{fl1984}.
This work was supported in part by the National Science Foundation through grant  PHY-1237510
and by the National Aeronautics and Space Administration through the University of Central Florida's NASA-Florida Space Grant Consortium.

%

\end{document}